\documentclass[superscriptaddress,twocolumn,amsmath,amssymb,nofootinbib]{revtex4-2}

\usepackage{bm,graphicx,color}
\usepackage{stmaryrd}
\usepackage{wasysym}
\newcommand{\doublehexagon}{{\varhexagon\!\varhexagon}}
\newcommand{\onehexmove}{\stackrel{\mbox{\tiny\varhexagon$\cdot\cdot$\varhexagon}}{\longleftrightarrow}}
\usepackage{amsthm}
\usepackage{enumitem}
\usepackage{xfrac}

\newtheorem{theorem}{Theorem}

\newcommand{\ket}[1]{\lvert#1\rangle}
\newcommand{\bra}[1]{\langle#1\rvert}

\begin{document}

\title{Quantum trimer models and topological SU(3) spin
liquids on the kagome lattice}
\author{Sven Jandura}
\affiliation{Ludwig-Maximilians-Universit\"at, Schellingstra\ss{}e 4, 80799 M\"unchen, Germany}
\affiliation{Institute for Theoretical Physics, ETH Z\"urich, Wolfgang-Pauli-Strasse 27, 8093 Z\"urich, Switzerland}
\affiliation{Max-Planck-Institute for Quantum Optics, Hans-Kopfermann-Stra\ss{}e 1, 85748 Garching, Germany}
\author{Mohsin Iqbal}
\affiliation{Max-Planck-Institute for Quantum Optics, Hans-Kopfermann-Stra\ss{}e 1, 85748 Garching, Germany}
\affiliation{Munich Center for Quantum Science and Technology, Schellingstra\ss{}e~4, 80799 M\"unchen, Germany}
\author{Norbert Schuch}
\affiliation{Max-Planck-Institute for Quantum Optics, Hans-Kopfermann-Stra\ss{}e 1, 85748 Garching, Germany}
\affiliation{Munich Center for Quantum Science and Technology, Schellingstra\ss{}e~4, 80799 M\"unchen, Germany}

\begin{abstract}
We construct and study quantum trimer models and resonating
$\mathrm{SU}(3)$-singlet models on the kagome lattice, which generalize
quantum dimer models and the Resonating Valence Bond wavefunctions to
a trimer and $\mathrm{SU}(3)$ setting. We demonstrate that these models
carry a $\mathbb Z_3$ symmetry which originates in the structure of
trimers and the $\mathrm{SU}(3)$ representation theory, and which becomes
the only symmetry under renormalization. Based on this, we construct simple
and exact parent Hamiltonians for the model which exhibit a topological $9$-fold
degenerate ground space. A combination of analytical reasoning and
numerical analysis reveals that the quantum order ultimately displayed by the model
depends on the relative weight assigned to different types of trimers -- 
it can display either $\mathbb Z_3$ topological order or form 
a symmetry-broken trimer crystal, and in addition possesses a 
point with an enhanced $\mathrm{U}(1)$ symmetry and critical behavior. Our results 
accordingly
hold for the $\mathrm{SU}(3)$ model, where the two natural choices for
trimer weights give rise to either a topological spin liquid or a system
with symmetry-broken order, respectively.  Our work thus demonstrates the
suitability of resonating trimer and $\mathrm{SU}(3)$-singlet ansatzes to
model $\mathrm{SU}(3)$ topological spin liquids on the kagome lattice.
\end{abstract}
\maketitle

\section{Introduction}

Spin liquids are exotic phases of matter where the competition between 
strong antiferromagnetic interactions and geometric frustration prevents
magnetic ordering, but instead gives rise to
topological order, that is, a global ordering in the structure of their
entanglement, and which thus display a range of exotic properties such as
fractional excitations with exotic
statistics~\cite{balents:spin-liquid-review-2010,savary:spin-liquids,knolle:spin-liquids}.
A paradigmatic model for topological spin liquids has been Anderson's
Resonating Valence Bond (RVB) state~\cite{anderson:rvb}. It is constructed
from different coverings of the lattice with $\mathrm{SU}(2)$
spin-$\tfrac12$ singlets, which are placed in a ``resonating''
superposition such as to further lower their energy.  Yet, the study of
RVB wavefunctions poses the challenge that different singlet coverings are
not orthogonal.  To alleviate this problem, quantum dimer models
have been studied instead, where inequivalent singlet configurations are
taken to be orthogonal, such as in a large-spin
limit~\cite{moessner:quantum-dimer-models}.

The study of dimer models reveals a strong dependence of their quantum
order on the type of lattice. Specifically, dimer models on bipartite
lattices are generally
critical~\cite{moessner:quantum-dimer-models,fradkin2013field}, while conversely on non-bipartite
lattices, in particular the triangular and kagome lattice, they 
have been found to exhibit $\mathbb Z_2$ topological
order~\cite{moessner:dimer-triangular,misguich:dimer-kagome}.  
Here, the case of the kagome lattice is of particular interest: First, the
dimer model on the kagome lattice is a renormalization (RG) fixed point
with a direct mapping to a $\mathbb Z_2$ loop
model~\cite{elser:rvb-arrow-representation} (i.e., Kitaev's Toric
Code~\cite{kitaev:toriccode}), and second, kagome Heisenberg
antiferromagnets are approximately realized in actual materials such as
Herbertsmithite~\cite{knolle:spin-liquids,norman:spin-liquid-review}, making the RVB state on the kagome lattice, despite not
being their exact ground state wavefunction, a particularly interesting
model to study.  By using Tensor Network methods, which allow
to construct smooth interpolations from the RVB to the dimer model and
which provide a powerful toolbox to directly probe for topological order, the
topological nature of the kagome RVB state could be unambiguously shown,
and the underlying parent Hamiltonians of the model were
identified~\cite{schuch:rvb-kagome,zhou:rvb-parent-onestar}.

Both the presence of $\mathbb Z_2$ topological order and the critical
physics found for bipartite lattices can be understood from the symmetries
of the system.
Dimer models naturally exhibit a $\mathbb Z_2$ gauge symmetry -- the
parity of the number of dimers leaving any given region is fixed, arising
from 
the fact that each dimer inside the region uses up two sites, or that two
spin-$\tfrac12$ make up a singlet -- suggestive of $\mathbb Z_2$
topological order. On the other hand, on bipartite lattices, the $\mathbb
Z_2$ is enhanced to a $\mathrm{U}(1)$ symmetry -- the number of $A$ and
$B$ sublattice dimers leaving a region must be equal, since each
(nearest-neighbor) dimer in
the region uses up an $A$ and a $B$ site -- indicative of critical
behavior. Thus, identifying the symmetries displayed by the dimer model
on different lattices is key to its understanding.

In recent years, $\mathrm{SU}(N)$ spin systems have received increasing
interest, in particular due to the possibility to engineer
$\mathrm{SU}(N)$-invariant interactions in the fundamental
representation in experiments with cold atomic
gases~\cite{gorshkov:suN-optical-lattice,scazza:suN-optical-lattice,zhang:suN-experiment}.
Hence, a natural first step is to consider 
$\mathrm{SU}(3)$-symmetric models in the fundamental representation. In that case,
singlets are tripartite, namely the fully antisymmetric state
$\sum_{ijk}\varepsilon_{ijk}\ket{i,j,k}$. Just as in the RVB construction, we
can think of building resonating $\mathrm{SU}(3)$ singlet states -- that
is, superpositions of different ways to cover the lattice with singlets --
to lower the energy, and just as for the RVB state, in the attempt
to analyze the model we are faced with the challenge that different
singlet configurations are lacking orthogonality. It is therefore natural
-- again in analogy to the RVB state -- to study quantum trimer models,
where different $\mathrm{SU}(3)$ singlets are replaced by orthogonal trimer
configurations. In this context, a few questions naturally arise:
First, what is the role of the underlying lattice -- for instance, is
there a similar dichotomy in the type of order displayed on bipartite vs.\ non-bipartite lattices?
Second, is there a ``natural'' lattice on which to define the trimer
model? This could either be a lattice on which the model forms an RG fixed
point with a direct mapping to a loop model, such as the kagome lattice
for the RVB state, or a lattice where the tripartite $\mathrm{SU}(3)$
singlets appear in a particularly natural way, which suggests a lattice
built from triangular simplices and thus yet again the kagome
lattice.

Previous work on trimer models and resonating $\mathrm{SU}(3)$ singlet
wavefunctions has been focused on the square lattice.  In particular, in
Ref.~\onlinecite{lee:resonating-trimer-state}, a quantum trimer model on
the square lattice has been introduced and it has been found 
that it is topologically ordered, and in
Ref.~\onlinecite{dong:su3-trimer-squarelattice}, a corresponding model
with resonating $\mathrm{SU}(3)$ singlets on the square lattice has been
identified as a $\mathbb Z_3$ topological spin liquid; both of these works
employed numerical analysis based on tensor networks. 
Indeed, such an order could have been expected, since trimers and
$\mathrm{SU}(3)$ singlets display a natural $\mathbb Z_3$ symmetry,
analogous to the $\mathbb Z_2$ symmetry in dimers.
On the other hand, in
Ref.~\onlinecite{kurecic:su3_sl} an $\mathrm{SU}(3)$ spin liquid on the kagome
lattice had been proposed whose ``orthogonal'' version
is an RG fixed point, namely the $\mathbb Z_3$ loop gas version of
the Toric Code, and which was shown to be topological. While this model
itself did not allow for an interpretation as a superposition of trimer patterns, a
modification of the $\mathrm{SU}(3)$ model which gave rise to such an
interpretation was also discussed, and numerically found to be in a
trivial phase.  
Together, these findings raise a number of questions: 
What is the nature of an orthogonal trimer model on the kagome lattice,
whose simplices naturally support trimers? 
Can we explain the different behavior seen for square vs.\ kagome lattice
from underlying symmetries in the model; in particular, are there
additional symmetries on top of $\mathbb Z_3$ emerging on the kagome
lattice which speak against topological order? 
And finally, can we obtain a more general understanding of the way in
which lattice geometries and the type of order in resonating
$\mathrm{SU}(N)$ wavefunctions could be related?

In this paper, we present a systematic analytical and numerical study of
the trimer model and its $\mathrm{SU}(3)$ variants on the kagome lattice.
Our key results are as follows:
\begin{itemize}
\item 
We present a simple mapping from the trimer model to a $\mathbb Z_3$ loop model (an ``arrow
representation'' similar to the kagome
RVB~\cite{elser:rvb-arrow-representation}), which however is missing
a vertex  configuration.  However, as we show analytically, 
all $\mathbb Z_3$-invariant loop configurations re-appear under blocking.
This shows that the model has the right symmetry to exhibit $\mathbb Z_3$
topological order, and that no additional symmetries are present
which would point to a different phase. This forms the starting point for
our subsequent analysis.
\item 
We demonstrate that both the trimer and the $\mathrm{SU}(3)$ model can exhibit
either $\mathbb Z_3$ topological order or conventional symmetry-breaking
order, as well as critical behavior.
Which one is realized is determined by the relative weight $\zeta$ of
different types of trimers (specifically, those on the elementary
simplices vs.\ the rest). There are two natural choices for this weight,
motivated by different interpretations of an equal weight superposition.
For the quantum trimer model, we find that it is topologically ordered for
both of these choices.  The resonating $\mathrm{SU}(3)$ singlet model, on
the other hand, is topologically ordered for one choice of $\zeta$ but  breaks
lattice symmetries for the other (the latter was observed in
Ref.~\onlinecite{kurecic:su3_sl}).  We further show that setting $\zeta$
to zero gives rise to an additional $\mathrm{U}(1)$ symmetry which
results in a critical
model.  
Finally, we provide phase diagrams for both the trimer and the
$\mathrm{SU}(3)$ model as a function of the weight $\zeta$, as well as a
detailed analysis
of the behavior as we interpolate between the trimer and the
$\mathrm{SU}(3)$ model.
\item 
We explicitly construct exact parent Hamiltonians for the trimer model. In the
arrow (i.e., loop model) representation, those Hamiltonians consist of
$3$-body terms across vertices and either $6$-body or $11$-body terms
which act around one or two
hexagons, respectively.  For the $11$-body Hamiltonians, we prove that
they give rise to the correct $9$-fold degenerate topological ground
space. Restricting to $6$-body terms (the same locality as for the RG
fixed point loop model) gives rise to $6$ additional ``frozen'' classical
ground states where the trimers display crystalline order, and which
do not couple to the remaining $9$ ground states; moreover, we show that
introducing a single $11$-body term anywhere in the lattice allows to melt
those crystals and recover the original $9$-fold degenerate ground space.
Decorating the trimers with $\mathrm{SU}(3)$ singlets increases the
locality of these terms to $8$ and $12$, respectively.
\end{itemize}

Together, our results show a remarkably rich behavior of the quantum
trimer and resonating $\mathrm{SU}(3)$-singlet models on the kagome
lattice, and refute a simple connection between the type of order and the
geometry of the lattice for trimer models analogous to the dimer case. Moreover, our construction
allows us to obtain a simple $\mathrm{SU}(3)$ spin liquid ansatz on the
kagome lattice which
has a natural interpretation as a resonating singlet pattern and thus as a
superposition of simple product states which are connected through local moves.
Finally, it highlights the importance of a new aspect in trimer models, and more
generally $N$-mer models, as opposed to dimer models, namely the key role
played by the relative weights assigned to different trimer
configurations.

The paper is structured as follows: In Section~\ref{sec:trimer-models}, we
introduce quantum trimer models, their arrow representation, and their
loop gas representation.  In Section~\ref{sec:z3-inj-ham}, we analyze the
symmetry of the trimer model in the loop picture and show that the full
$\mathbb Z_3$ symmetry (technically, $\mathbb
Z_3$-injectivity~\cite{schuch:peps-sym}) is
restored under blocking, which proves the absence of additional symmetries
and implies the existence of local parent Hamiltonians.  We then proceed
to construct the simplest such parent Hamiltonian, with the technical
arguments given in two appendices.   In
Section~\ref{sec:order-and-phasediag}, we combine analytical reasoning
with numerical tensor network methods to investigate the type of order
which the quantum trimer model exhibits, and in particular how the nature
of the phase -- topological, symmetry broken, or critical -- depends on
the relative weight of different trimers.  Finally, in
Section~\ref{sec:su3-model} we generalize the trimer model by equipping it
with $\mathrm{SU}(3)$ singlets, which we then continuously connect to a
pure resonating $\mathrm{SU}(3)$-singlet model with the fundamental
representation acting on each site.  We study the phase
diagram of the $\mathrm{SU}(3)$ model as well as the physics along the
trimer $\leftrightarrow$ $\mathrm{SU}(3)$  interpolation, and discuss the corresponding
parent Hamiltonian. The section closes with a brief discussion of the
$\mathrm{SU}(3)$ model as a variational ansatz.

\section{Trimer models\label{sec:trimer-models}}

\begin{figure}
\includegraphics[width=\columnwidth]{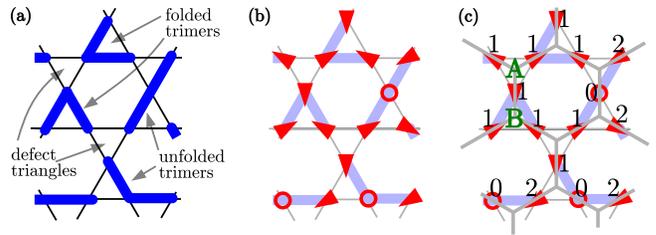}
\caption{
\textbf{(a)} A trimer covering $T$. Folded trimers can be either
treated as different depending on their orientation (as the two shown), or
as identical (i.e.\ interpreted as a set of three vertices). We
take the latter approach but allow to assign a different weight $\zeta$ to
folded trimers, which allows to resolve them internally as a superposition
of inequivalent trimers. \textbf{(b)} Arrow representation: We assign
arrows to the outer vertices, pointing into the triangle towards the
center of the trimer, $\circ$ to inner vertices, and three inpointing
arrows to folded trimers. This establishes a one-to-one mapping from
trimer coverings to arrow patterns which obey a Gauss law, where the
configuration with three $\circ$ is forbidden.  \textbf{(c)} Loop
representation. Arrows are replaced by $1$ and $2$, respectively,
depending whether they point from the $A$ to the $B$ sublattice or vice
versa, and $\circ$ by $0$.  The loop representation again satisfies a
Gauss law with $000$ forbidden.}
\label{fig:trimer-covering}
\end{figure}

Let us start by defining trimer models, Fig.~\ref{fig:trimer-covering}a. A 
trimer on the kagome lattice is a
covering of two adjacent edges and correspondingly three vertices. For a
trimer, we distinguish two \emph{outer} and one \emph{inner} vertex, and we
distinguish \emph{folded} trimers (living on a single triangle) from
\emph{unfolded}
ones (living on two triangles).  A trimer covering $T$
is a complete covering of the lattice
with non-overlapping trimers, i.e., each vertex is contained in exactly
one trimer. We can treat the trimer coverings $T$ as an orthonormal basis
$\ket{T}$ of a Hilbert space; an \emph{(orthogonal) trimer model} is then given as the
equal-weight superposition of all trimer configurations $\mathcal T$, 
\begin{equation}
\label{eq:trimerstate}
\ket{\Psi_\mathrm{trimer}} = \sum_{T\in \mathcal T} \ket{T}\ .
\end{equation}
A key question will be how to treat folded trimers: Do we think of them as
three different trimers (related by rotation), or do we treat them all as
the same trimer (considering a trimer rather as an unordered set of three
vertices), see Fig.~\ref{fig:trimer-covering}a?  Here, we will treat
folded trimers as equivalent (i.e.\ as a set of vertices), but we will
allow to assign them a different weight $\zeta$ [cf.\
Eq.~\eqref{eq:trimerstate-loop}], accounting for multiple ``internal''
degrees of freedom such as differently oriented trimers (e.g.\
$\zeta=\sqrt{3}$ would allow to resolve folded trimers in three orthogonal
ways with weight $1$ each).

A local representation of $\ket{T}$ can be constructed by assigning
either arrows $\{\blacktriangleright,\blacktriangleleft\}$ or a no-arrow
symbol $\circ$ to every vertex (see Fig.~\ref{fig:trimer-covering}b): To
each outer vertex of a trimer, 
we assign an arrow pointing into the triangle to which
the trimer extends, and to inner
vertices, we assign $\circ$. The exception are folded trimers, 
in which case we assign inpointing arrows 
to all three vertices. This associates a unique arrow pattern to each trimer
covering. Conversely, any arrow pattern for which
(\emph{i})~the number of inpointing minus outpointing arrows is $0\ \mathrm{mod}\:3$
and (\emph{ii})~triangles with three $\circ$ are forbidden, maps
uniquely to a trimer pattern.

Let us now consider the arrow degrees of freedom as link variables
on the dual honeycomb lattice. 
We replace the arrows by $1$ and $2\equiv -1$ (we work
$\mathrm{mod}\, 3$ from now on), depending on whether they point from
the $A\blacktriangleright B$ sublattice of the honeycomb  or vice
versa; $\circ$ will be replaced by $0$. The arrows then correspond to
configurations with a $\mathbb Z_3$ Gauss law around each vertex of the
honeycomb lattice, where the configuration $000$ is forbidden. In this
language, the trimer model corresponds to 
\begin{equation}
\label{eq:trimerstate-loop}
\ket{\Psi_\mathrm{trimer}} = 
	\sum_{L\in \mathcal L}
	\zeta^{N^A_{222}+N_{111}^B} \ket{L}\ ,
\end{equation}
where the $\mathbb Z_3$ edge configurations $L$ on the honeycomb lattice 
are taken from the set $\mathcal L$ of all Gauss law configurations
(``loop configurations'') with no $000$ around any vertex; here, $\zeta$
controls the relative weight of folded trimers (which correspond to
$222$ configurations around $A$ vertices and $111$ around $B$ vertices,
counted by $N^A_{222}$ and $N^B_{111}$).

\begin{figure}
\includegraphics[width=8cm]{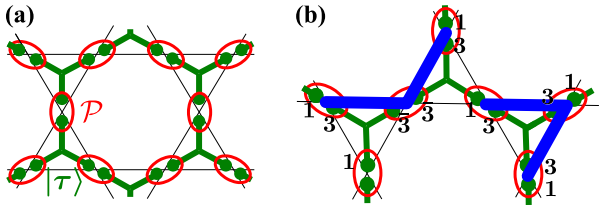}
\caption{%
\textbf{(a)} PEPS construction for the trimer and $\mathrm{SU}(3)$ model:
The wavefunction is constructed by starting with fiducial states
$\ket\tau$ which enforce the Gauss law, and subsequently applying maps
$\mathcal P$ to them which output the physical degrees of freedom on
the kagome lattice.
\textbf{(b)}
Construction of a trimer model with $\mathrm{SU}(3)$ degrees of freedom:
The fiducial states transform as a singlet in $(\bm 1\oplus\bm 3 \oplus
\bar{\bm 3})^{\otimes 3}$, where arrows point from $\bm 1$ to $\bm 3$ and
thus the singlet in 
$\bar{\bm 3}\otimes \bar{\bm 3}\otimes \bar{\bm 3}$
is forbidden; it is thus either a $\bm 3\otimes \bm 3\otimes \bm 3$
singlet or a singlet in one of the pairs $\bm 3 \otimes \bar{\bm
3}\otimes\bm 1$.
The resulting state is a trimer model decorated with
$\mathrm{SU}(3)$ degrees of freedom.  By projecting onto the $\bm 3$ in 
$\bar{\bm 3}\otimes \bar{\bm 3}=\bm 3 \oplus \bar{\bm 6}$, trimers are
decorates with $\mathrm{SU}(3)$ singlets.
}
\label{fig:peps}
\end{figure}

To study the properties of this model it is convenient to use a tensor
network or PEPS (Projected Entangled Pair State) representation, which
provides us with a range of powerful analytical and numerical tools for
its
analysis~\cite{verstraete:mbc-peps,verstraete:2D-dmrg,bridgeman:interpretive-dance,perez-garcia:parent-ham-2d,schuch:peps-sym,haegeman:shadows,duivenvoorden:anyon-condensation}.
The corresponding PEPS is constructed from two types of tensors,
see Fig.~\ref{fig:peps}a,
similar to e.g.\ the tensor network for the RVB and dimer
model~\cite{schuch:rvb-kagome}:
 One type
of tensor -- corresponding to a state $\ket\tau=\sum {t_{ijk}}\ket{i,j,k}$
-- only has virtual indices and sits inside triangles, it ensures only
valid vertex configurations appear (with weight $\zeta$ assigned to
on-site triangles), that is,
$t_{ijk}=|\varepsilon_{ijk}|+\delta_{i=j=k=1}+\zeta\delta_{i=j=k=2}$. 
The other tensor -- corresponding to a map $\mathcal P=\sum
P^a_{ij}\ket{a}\bra{i,j}$ -- sits on the
edges and maps two virtual degrees of freedom, one from each of the
adjacent tensors, to a physical arrow state $\ket{a}$, that is,
$P^\blacktriangleright_{12}=P^\blacktriangleleft_{21}=P^\circ_{00}=1$ and
zero otherwise, with the arrow oriented towards the $2$.  The trimer
wavefunction $\ket{\Psi_\mathrm{trimer}}$ is then obtained by arranging
the tensors on the honeycomb lattice and contracting the virtual indices,
or alternatively applying the maps $\mathcal P$ to the states $\ket\tau$,
see Fig.~\ref{fig:peps}a.   (Note that on $B$ triangles, the role of $1$ and
$2$ is swapped relative to Fig.~\ref{fig:trimer-covering}c; this can be
changed by an appropriate gauge transformation.)

\section{$\mathbb Z_3$-injectivity and parent
Hamiltonian\label{sec:z3-inj-ham}}

We will now address the question whether the model is topologically
ordered, and whether it appears as a ground state of a local parent
Hamiltonian with a suitable (topological) ground space structure. To start
with, the modified
model which is an equal weight superposition of \emph{all} Gauss law patterns --
where we include the $000$ configuration and let $\zeta=1$ -- is a
topological RG fixed point model, namely the $\mathbb Z_3$ version of
Kitaev's toric code model~\cite{kitaev:toriccode}. But how severe is the
exclusion of the $000$ configuration?  Does it induce a stronger symmetry
than the $\mathbb Z_3$ Gauss law -- e.g., a $\mathrm{U}(1)$ conservation
law which would be indicative of a critical phase -- or does it merely induce a finite
length scale without breaking the topological order?

\begin{figure}
\includegraphics[width=\columnwidth]{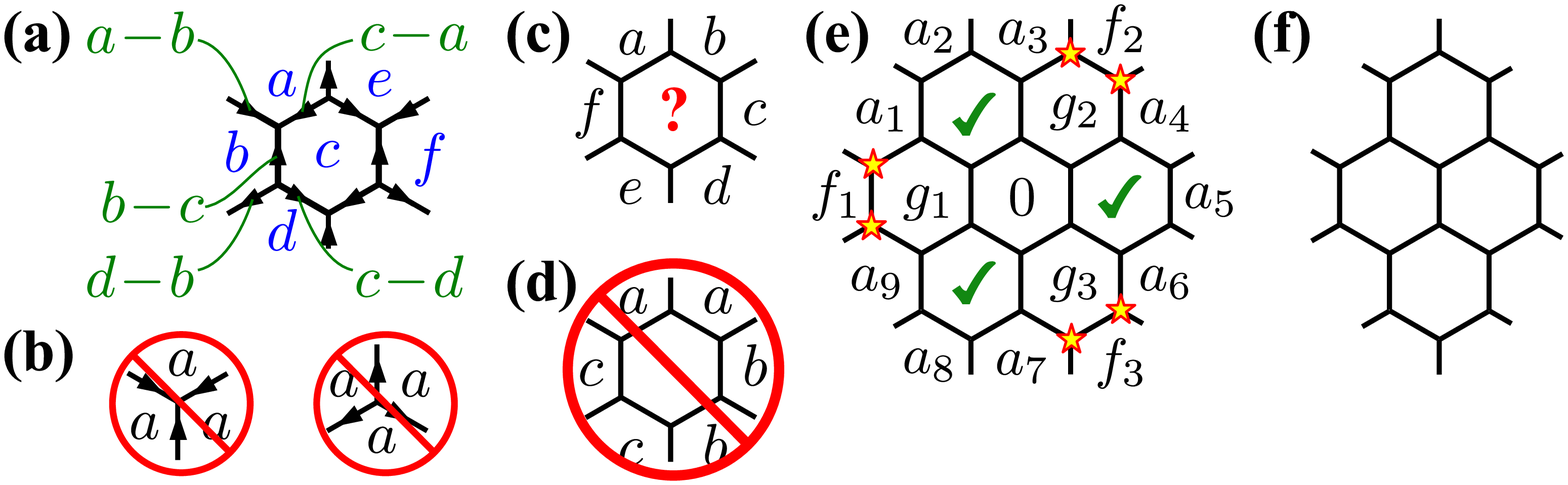}
\caption{
\textbf{(a)} Height representation of the trimer model: A
$\mathbb Z_3$ ``height variable'' (blue labels) is associated to every plaquette
such that link variables (green labels) are the (oriented) differences of the
height variables. \textbf{(b)} Forbidden height configurations
(corresponding to the forbidden $000$ configurations of the edges).
\textbf{(c,d)} The central plaquette in (c) can be assigned a valid value
exactly if the boundary is not in the state (d) or rotations thereof, with
$a,b,c$ all different. \textbf{(e)} Region used for the $\mathbb
Z_3$-injectivity proof (see text). \textbf{(f)} Smallest $\mathbb
Z_3$-injective region, i.e.\ where all boundary configurations are
admissible.}
\label{fig:injective}
\end{figure}

To understand this, we study what happens as we block sites: Does the
missing $000$ configuration result in missing configurations at all length
scales -- which would suggest additional conservation laws -- or is the
full symmetry restored?  To this end, we consider a dual ``height''
representation of the model, where we assign $\mathbb Z_3$ variables to
the plaquettes (Fig.~\ref{fig:injective}a, where the plaquette variables
are labeled by roman letters), such that the edge degrees of freedom are
obtained as the difference of plaquette variables (oriented
clockwise/counter-clockwise around $B$/$A$ sublattice sites).  This
mapping from plaquettes to edges is $3$-to-$1$, where the forbidden $000$
configuration rules out three identical plaquette variables around a
vertex (Fig.~\ref{fig:injective}b).
Now consider first the neighborhood of one hexagon shown in
Fig.~\ref{fig:injective}c: Given labels $a,\dots,f$ at the boundary, when
can we assign an allowed label to the central plaquette?  It is easy to
see that this is the case if and only if the pattern is not of the form
Fig.~\ref{fig:injective}d for $a,b,c$ all different, or rotations thereof.  Now consider
the block in Fig.~\ref{fig:injective}e, with arbitrary plaquette variables
$\{a_i\}$ and $\{f_j\}$ assigned to the boundary -- corresponding to an
arbitrary boundary configuration of the edges which satisfies the $\mathbb Z_3$
Gauss law.  Next,  for $f_i=0,1,2$ assign $g_i=1,2,1$.
This way, $f_i\ne g_i$ (and
thus the condition Fig.~\ref{fig:injective}b is satisfied around all
vertices marked $\star$), and
$g_i\ne 0$. Finally, we assign $0$ to the central plaquette. 
We now immediately see that 
we cannot have the pattern of Fig.~\ref{fig:injective}d
around the plaquettes marked with green tickmarks, 
and thus, we can also assign
consistent value to these.
  That is, any $\mathbb
Z_3$-invariant loop configuration at the boundary has a realization on the
interior, and is thus an admissible boundary configuration: We thus find
that the $\mathbb Z_3$-invariant space is restored after blocking, proving
that removing the $000$ vertices from the $\mathbb Z_3$ loop model 
induces no additional constraints under renormalization.  In fact, this
result still holds for the smaller patch in Fig.~\ref{fig:injective}f, as
can be verified by an
exhaustive search.

The fact that the $\mathbb Z_3$-invariant subspace is restored under
blocking -- or, in the language of tensor networks, $\mathbb
Z_3$--injectivity~\cite{schuch:peps-sym} is reached (that is, the blocked tensor describes an
injective map from boundary to bulk on the $\mathbb Z_3$--invariant
subspace) implies the existence of a local parent Hamiltonian with a
$9$-fold degenerate ground space on the torus, which is spanned by the
trimer state $\ket{\Psi_\mathrm{trimer}}$ and its topologically equivalent
siblings, obtained by assigning a phase $\omega^{\nu_hN_h+\nu_vN_v}$
($\omega=e^{2\pi i/3}$) with $\nu_h,\nu_v=0,1,2$ to
configurations for which the link variables sum to $N_h$ ($N_v$)
along a horizontal (vertical) loop around the
torus~\cite{schuch:peps-sym,schuch:rvb-kagome}. Note, however, that
this does not necessarily imply that the model is topologically ordered in
the thermodynamic limit (ground states can vanish, or additional low
energy state can appear)~\cite{schuch:topo-top}, and numerical study is required 
in addition to
unambigously assess the topological nature of the trimer model; we will
turn to this in the next section.

For now, let us discuss the form of the parent Hamiltonian. Generally,
the local terms in the parent Hamiltonian are positive semi-definite
operators (e.g.\ projectors) which are constructed such that they are zero
exactly on all allowed states on the spins supporting them (in a tensor
network, this is the space spanned by choosing arbitrary boundary
conditions)~\cite{perez-garcia:parent-ham-2d,schuch:peps-sym}. For the case of loop-like models
like the one at hand, one way to explicitly construct such Hamiltonians is
to build them from two types of terms: The first consists of $3$-body
projectors which act across vertices and have precisely the allowed vertex
configurations in their kernel; here, those are the $\mathbb Z_3$
Gauss law configurations except $000$.  The second type of terms couples
different loop configurations through local transitions (that is, its
ground space is spanned by the superposition $\ket{\chi_k}$ of coupled loop
configurations with the correct relative weights, $h=\openone-\sum
\ket{\chi_k}\bra{\chi_k}$), in such a way that any
two loop configurations in the same topological sector (distinguished by
their winding number, i.e., dual to the basis above) can be coupled
through a sequence of such local moves induced by the individual
Hamiltonian terms.

On what region do the latter type of Hamiltonian terms have to act?  Using
the $\mathbb Z_3$-injectivity after blocking, such regions can be
constructed using canonical techniques; in essence, they need to contain
an injective region plus a thin surrounding in a way which allows patching
regions together%
~\cite{perez-garcia:parent-ham-2d,schuch:peps-sym,schuch:rvb-kagome,molnar:normal-peps-fundamentalthm}.
Given the minimum injective region identified in Fig.~\ref{fig:injective}f,
this gives terms acting on finite but still rather large regions.
However, as we show in Appendix~\ref{sec:app:2hexham},  one can do much better: In
order for the Hamiltonian 
to couple all configurations in a topological sector, it is sufficient to
have Hamiltonian terms which act on \emph{two adjacent hexagons}, or $11$
edge degrees of freedom in the arrow or loop representation! 

We thus find that in order to ensure a topologically degenerate ground
space, one needs a Hamiltonian of the form
\begin{equation}
\label{eq:ham-doublehex}
H =  \sum h_{\Yleft} + \sum h_{\doublehexagon}
\end{equation}
where $h_{\Yleft}$ is a projector acting on vertices which has the allowed
vertex configurations as their ground space, and
$h_{\doublehexagon}$ is a
projector acting on pairs of adjacent plaquettes which, for every choice
of surrounding degrees of freedom, has the equal
weight superpositions of all allowed configurations consistent with those
boundaries as their ground
space;\footnote{Note that the latter term needs no access to the
surrounding degrees of freedom: Their value is unchanged and can be
inferred from the degrees of freedom in the two hexagons using Gauss' law.} the sums run
over all vertices and pairs of adjacent hexagons, respectively.

\begin{figure}
\includegraphics[width=8cm]{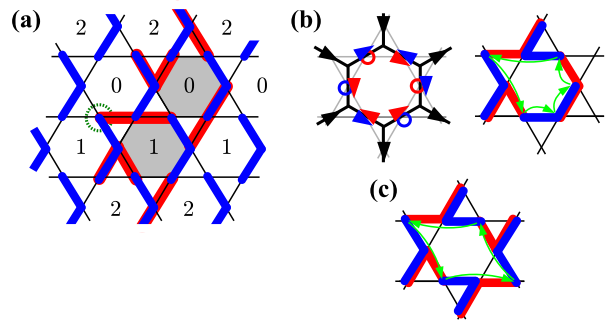}
\caption{%
\textbf{(a)}
``Frozen'' trimer configuration (blue) with crystalline order which cannot be melted by
one-hexagon moves alone, and thus forms an independent ground state under
one-hexagon parent Hamiltonians; this can be seen immediately from
the height representation (labels inside hexagons). A two-hexagon move is
needed to start melting the crystal, illustrated by the red trimer
configuration which requires updating the two hexagons marked gray. 
\textbf{(b)} When equipping the trimers with $\mathrm{SU}(3)$ singlets,
a Hamiltonian which acts on one hexagon in the arrow representation (left) also
might have to act on some outer vertices due to the entanglement of the trimer
(right).  Here, the Hamiltonians acts by coupling the blue and red
configurations.
\textbf{(c)} For single-hexagon moves, in the worst case it is required to
act on 
two degrees of freedom outside of the hexagon.  Similarly, for the
two-hexagon move in (a), one has to act on at least one additional outside
degree of freedom, as the one marked by the dashed green circle.
}
\label{fig:hamiltonian}
\end{figure}

What happens when we go even further and restrict to one-hexagon
Hamiltonians, $H=\sum h_{\Yleft}+\sum h_{\varhexagon}$,  where
$h_{\varhexagon}$ induces transitions between all allowed configurations on
a hexagon (which is the same locality as for the $\mathbb Z_3$ fixed point
loop model)?  It turns out that in that case, additional ground states
appear. Those states are all of the form shown in
Fig.~\ref{fig:hamiltonian}a, that is, they exhibit crystalline
order.\footnote{There are $6$ such states related by symmetry.} Using the
height representation (see figure), it can be easily seen
that it is impossible to change the value of a single plaquette without
violating Fig.~\ref{fig:injective}b, and an update acting on two hexagons
is needed to start melting the crystal (shown red in
Fig.~\ref{fig:hamiltonian}a). 

As it turns out -- discussed in detail in Appendix~\ref{sec:app:1hexham}
-- these are the \emph{only} additional ground states
which appear when restricting to one-hexagon terms, while all other
configurations in each topological sector can be coupled even by
one-hexagon updates. Differently speaking, crystals of the form
Fig.~\ref{fig:hamiltonian}a which only cover part of the system can
be melted from the outside, and in particular, it is possible to change
the configuration anywhere inside the crystal by melting a channel through
the crystal, updating the configuration as needed, and re-freezing it
towards the outside.  This also implies that \emph{a single}
two-hexagon term
$h_{\doublehexagon}$
\emph{anywhere} in the system is sufficient to couple the crystalline
configurations to the others, and this way get back a Hamiltonian with $9$ ground
states.\footnote{Note, however, that in this case configurations which
only differ in one location are typically connected by a path of updates
with a length on 
the order of the system size $N$, that is, in $N$'th order,
which suggests that the corresponding system will be gapless.}

\section{Order and phase diagram\label{sec:order-and-phasediag}}

Is the trimer model on the kagome lattice topologically ordered, and 
does this depend on the weight $\zeta$ of folded trimers? 
To this end, let us first see what we can understand analytically,
e.g.\ in limiting cases, about the effect of varying $\zeta$. 
To start with, observe that any unfolded trimer uses up two triangles per three
vertices.  Folded trimers, on the other hand, only require one triangle
per three vertices. As the kagome lattice has $2$ triangles per $3$
vertices, each folded trimer is therefore accompanied by exactly one \emph{defect
triangle}, that is, a triangle with no edge of a trimer on it
(Fig.~\ref{fig:trimer-covering}a). Thus, we find that the trimer state
\eqref{eq:trimerstate-loop} is a sum over all trimer configurations, where
configurations with each $K$ folded trimers and $K$ defect triangles  
are weighted by
$\zeta^K$. (Curiously, this implies that only the product of the reweighting of 
folded trimers and defect triangles matters). From this, we can understand
two limiting cases: For $\zeta\to\infty$, the system consists solely of
folded trimers and defect triangles, and
will thus break the lattice symmetry by putting all (folded) trimers
either only on up- or only on down-pointing triangles; such a type of crystalline order
has indeed been identified for the ground state of the $\mathrm{SU}(3)$
Heisenberg model on the kagome lattice (see also
Section~\ref{sec:su3-model}). On the other hand, for
$\zeta=0$, folded trimers and defect triangles disappear, and thus the
configurations $111$ and $222$ are forbidden (in addition to the already forbidden
$000$).   The remaining configurations are $012$ and its permutations,
i.e., one in- and one outpointing arrow at every vertex.  At 
$\zeta=0$, the 
system thus possesses a $\mathrm{U}(1)$ symmetry, suggestive of critial
behavior (as one can construct a $\mathrm{U}(1)$ height representation and
thus an effective $\mathrm{U}(1)$ field theory description).

In order to understand the behavior of the system also away from those
limiting cases,  we use tensor network techniques introduced in earlier
work to study the nature of the trimer
model~\cite{duivenvoorden:anyon-condensation,iqbal:z4-phasetrans,kurecic:su3_sl,iqbal:rvb-perturb,iqbal:breathing-kagome}.
Specifically, we compute iMPS fixed points of the transfer matrix from
left and right, and use their symmetry breaking pattern with respect to
the $\mathbb Z_3\times \mathbb Z_3$ symmetry in ket+bra to
identify the topological nature of the system; this method also allows us
to extract correlation lengths $\xi$ for the trivial and anyon-anyon
correlators (where the latter correspond to the inverse anyon mass).  

\begin{figure}
\includegraphics[width=\columnwidth]{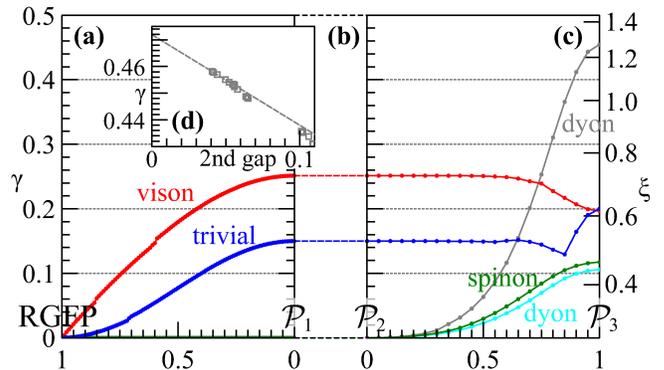}
\caption{
Length scales $\xi$  for two-point correlations (labeled ``trivial'') and
anyon-anyon correlators (i.e.\ inverse anyon masses, labeled by anyon
type), along a sequence of interpolations. 
The $y$ axis is $\gamma=e^{-1/\xi}$.
\textbf{(a)} Interpolation from
the loop model
(RG fixed point) to the trimer model, by removing the $000$ configuration.
No spinon correlations appear, as different trimers configurations remain
orthogonal.  \textbf{(b,c)} Trimers are equipped with an $\mathrm{SU}(3)$
representation, see text: Interpolation (b) removes the $\bar{\bm 6}$ in the center
of the trimer and has no effect, and yields trimers equipped with
$\mathrm{SU}(3)$ singlets at the point $\mathcal P_2$. Interpolation (c)
removes the arrow information; this gives rise also to spinon and dyon
correlations as orthogonality of trimer configurations is lost.
However, correlations remain finite all the way to the $\mathrm{SU}(3)$
point $\mathcal P_3$. \textbf{(d)}~Inset: Extrapolation of the leading
correlation (dyon mass) at the $\mathrm{SU}(3)$ point for increasing iMPS
bond dimension, using the
$\varepsilon$-$\delta$-method~\cite{rams:epsilon-delta-extrapol}.
}
\label{fig:interpol}
\end{figure}

First, we study an interpolation from the RG fixed point to the trimer
model at $\zeta=1$, obtained by decreasing the weight of $000$
configurations. This corresponds to a smooth interpolation of parent
Hamiltonians (as $\mathbb Z_3$-injectivity is
kept)~\cite{schuch:rvb-kagome}, and is thus a
reliable way to certify the absence of phase transitions, in addition to
the symmetry breaking pattern of the topological $\mathbb Z_3\times
\mathbb Z_3$ symmetry in
the entanglement.  The result is shown in Fig.~\ref{fig:interpol}a, where
the $x$ axis gives the weight of the $000$ configuration in
the superposition.   We see that as we decrease the weight of the $000$ configuration,
correlations in the system build up, up to
$\xi\approx0.72$ for the trimer point. 
 The dominant length scale is given by visons, while
spinons or combined vison-spinon (``dyon'') correlations remain zero. This
is to be expected, as visons correspond to a disbalance in different
Gauss law (i.e., loop) configurations (induced by suppressing $000$), while
spinons correspond to breaking up trimers (violations of the Gauss law),
which doesn't occur as different trimers remain orthogonal.\footnote{In
the PEPS picture, vison pairs correspond to strings of $\mathbb Z_3$
symmetry actions, while spinons correspond to objects which transform as a
non-trivial irrep under the symmetry, placed on the links when contracting
tensors~\cite{schuch:rvb-kagome,kurecic:su3_sl,iqbal:rvb-perturb,iqbal:breathing-kagome}.}

\begin{figure}
\includegraphics{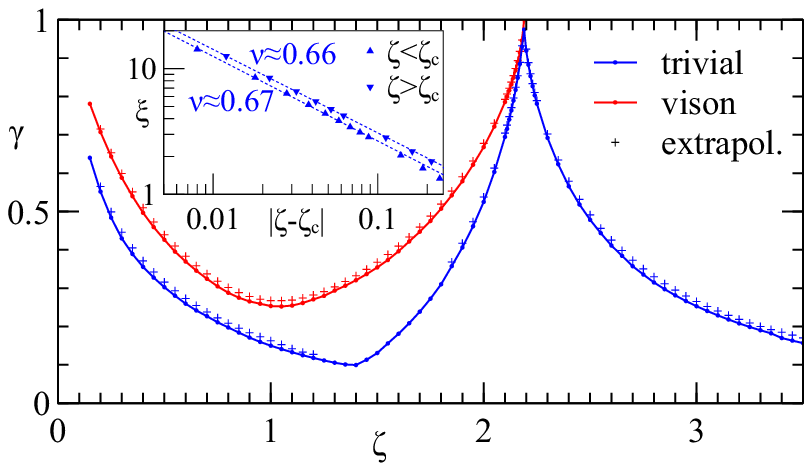}
\caption{
Correlation length $\xi$ vs.\ folded trimer weight $\zeta$ for the
trimer model. We plot $\gamma=e^{-1/\xi}$ for trivial and vison
correlations. We identify a $\mathbb Z_3$ topological phase for
$\zeta<\zeta_c\approx2.19$, a $\mathbb Z_2$ symmetry breaking phase for
$\zeta>\zeta_c$,
and a critical point at/around $\zeta=0$. Data shown is for an
iMPS truncation error of $\eta=10^{-4}2^{-38}\approx3.64\times10^{-16}$, crosses give extrapolations obtained
with the $\varepsilon$-$\delta$-method~\cite{rams:epsilon-delta-extrapol}
(where reliable). The inset shows the critical scaling around
$\zeta_c=2.188$ for the extrapolated values, which yields a critical
exponent $\nu\approx2/3$ at both sides of the transition.
}
\label{fig:zeta-diag}
\end{figure}

Let us next consider the phase diagram as a function of the weight $\zeta$
of folded trimers. Our results are shown in Fig.~\ref{fig:zeta-diag},
where we plot trivial and (in the topological phase) vison correlations vs.\ $\zeta$.
Together with the analysis of the ordering relative to the $\mathbb Z_3$
symmetry, we
indentify a $\mathbb Z_3$ topological phase around $\zeta=1$, where we had
already established topological order by interpolation from the RG fixed
point. Around $\zeta_c\approx2.19$, we observe a transition into a $\mathbb
Z_2$ symmetry breaking phase -- this is exactly the phase we discussed
above, where for $\zeta\to\infty$, folded trimers and defect triangles
alternate.  On the other hand, for $\zeta\to0$, the correlations diverge,
consistent with critical behavior induced by the $\mathrm{U}(1)$ symmetry
at $\zeta=0$ discussed above. 
Noteworthily, $\zeta_c>\sqrt{3}$, i.e., the trimer
model is topologically ordered independent of whether we interpret folded trimers 
as a single object
or as a sum of three orthogonal trimer realizations.  The inset shows the
scaling behavior of the correlation length $\xi$ for a critical point
$\zeta_c=2.188$, which
yields a critical exponent $\nu\approx2/3$. Remarkably, this is consistent with
the $4$-state~(!) Potts transition.

\section{SU(3) model\label{sec:su3-model}}

It is natural to use the trimer model to build an $\mathrm{SU}(3)$
model. If we attach the fundamental $\bm{3}$ representation to each
vertex, an $\mathrm{SU}(3)$ singlet consists of three spins
$\bm3\otimes\bm3\otimes\bm3$ and is of the form $\sum
\varepsilon_{ijk}\ket{i,j,k}$ (with $\varepsilon$ the fully antisymmetric
tensor). By replacing each trimer with an
$\mathrm{SU}(3)$ singlet (suitably oriented), we thus arrive at an
$\mathrm{SU}(3)$ resonating trimer state, in analogy to $\mathrm{SU}(2)$
resonating valence bond states.

Is the physics of the model affected by replacing the trimers by
$\mathrm{SU}(3)$ singlets? Since unlike abstract trimers, different singlet
configurations are not orthogonal, this is far from obvious. In order to
assess this question, we employ a tensor network representation of the
model which allows to treat the (orthogonal) trimer model and the
$\mathrm{SU}(3)$ model on the same footing; it can be seen as a variant of
the $\mathrm{SU}(3)$ model in Ref.~\onlinecite{kurecic:su3_sl}, see
Fig.~\ref{fig:peps}a.
 We start by triangular states
$\ket\tau$ of three sites with representation $\bm 1 \oplus \bm 3 \oplus
\bar{\bm{3}}$ each, where $\ket\tau$ is an equal weight superposition of
the $8$ possible singlets without the one in $\bar {\bm 3}\otimes \bar{\bm
3} \otimes
\bar{\bm 3}$, and where we choose the amplitudes of all singlets $+1$, except
for the one in $\bm 3\otimes \bm 3\otimes \bm 3$, which has amplitude 
$i\,\zeta$; this way, $\ket\tau$ is rotational invariant and transforms
under reflection $\mathcal R$ as
$\mathcal R\,\ket\tau = \ket{\bar\tau}$. Next, we place $\ket\tau$ on the simplices
of the kagome lattice and apply maps $\mathcal P(\vec\theta)$ which depend on
some interpolation parameters $\vec\theta$. There are three special
points for 
$\mathcal P$ between which we interpolate: $\mathcal P_1$ projects each site onto $(\bm 1\otimes \bm 3)
\oplus (\bm 3 \otimes \bm 1) \oplus (\bar{\bm3}\otimes\bar{\bm3})$ --
  this
is unitarily equivalent to the (orthogonal) trimer model, by associating
$\bm 1\otimes \bm 3$ to $\blacktriangleright$, and
$\bar{\bm3}\otimes\bar{\bm3}$ to $\circ$. 
(Gauss law is ensured since $\ket\tau$ 
is a singlet.)
 This assignment produces a direct correspondence
between
configurations of $\ket\tau$ and trimer patterns, cf.\
Fig.~\ref{fig:peps}b. 
Next, $\mathcal P_2$ is obtained from
$\mathcal P_1$ 
by removing the $\bar{\bm6}$ from
the subspace $\bar{\bm3}\otimes\bar{\bm3}=\bm{3}\oplus\bar{\bm6}$ while
keeping the total weight of the subspace (i.e., reweighting it by
$\sqrt{3}$ -- not doing so would suppress $\bar{\bm3}\otimes
\bar{\bm 3}$ and thus unfolded trimers). After removing the
$\bar{\bm 6}$, the effective Hilbert space is $\bm 3\oplus \bm 3\oplus \bm
3 \cong\bm 3 \otimes \mathbb C^3$. Finally, $\mathcal P_3$ removes the
degeneracy space $\mathbb C^3$ by projecting on the equal weight
superposition of the three $\bm 3$ (where the relative phases $\pm1$ are
chosen such that $\mathcal P_3\mathcal R =-\mathcal P_3$, which
yields a rotationally invariant 
wavefunction $\ket\Psi$ which transform as $\mathcal R\ket\Psi =
\ket{\bar\Psi}$). The interpolation $\mathcal
P_1\to\mathcal P_2\to\mathcal P_3$ can be carried out continuously -- where
along the entire $\mathcal P_1\to\mathcal P_2$ interpolation, we keep the weight
of the $\bar{\bm 3}\otimes\bar{\bm 3}$ subspace constant --
allowing us to go smoothly from the orthogonal trimer model to the
$\mathrm{SU}(3)$ model.\footnote{In Fig.~\ref{fig:interpol}, we plot the 
data vs.\ the amplitude which we change
along the interpolation.}
We refer the reader to
Ref.~\onlinecite{kurecic:su3_sl} for further details.

The behavior along the interpolation is shown in
Fig.~\ref{fig:interpol}bc. Notably, interpolating $\mathcal P_1\to\mathcal
P_2$ does not induce any change in the wavefunction due to the
reweighting. This can be understood since $\bar{\bm3}\otimes\bar{\bm3}$ only appears
in the middle of unfolded trimers, and different trimers remain orthogonal
-- the interpolation is thus merely a local basis transformation,
while without the reweighting, unfolded trimers would have been
suppressed.\footnote{Along the interpolation $\mathcal P_1\to\mathcal P_2$, the entangled state put on
top of the trimer is -- up to normalization -- of the form $(1\!\!1\otimes
W_\theta\otimes 1\!\!1)(\ket{\omega}\otimes\ket{\bar\omega})$, where
$\ket\omega$ is the singlet in $\bm 3\otimes \bar{\bm 3}$,
$\ket{\bar\omega}$ the reflected version of $\ket\omega$, and $W_{\theta}$
decreases the weight of $\bar{\bm 6}$ in $\bar{\bm 3}\otimes\bar{\bm
3}=\bm 3\oplus \bar{\bm 6}$ all the way to zero (which gives the
antisymmetric state).\label{foot:trimer-state}}
When
interpolating $\mathcal P_2\to\mathcal P_3$, orthogonality of 
different trimer configurations
is lost, which induces a finite length scale also
for spinons and combined dyonic excitations; we find that for the
$\mathrm{SU}(3)$ point, the dominant length scale is a dyonic one, with
$\xi\approx1.3$. Importantly, from Fig.~\ref{fig:interpol}, together
with an extrapolation of the correlations, it
is clear that the correlations remain finite, and the $\mathrm{SU}(3)$
resonating trimer model is in the $\mathbb Z_3$ topological phase;
for the $\mathrm{SU}(3)$ point, the extrapolation (Fig.~\ref{fig:interpol}d) 
yields a dominant dyon correlation length of $\xi=1.33$.

\begin{figure} 
\includegraphics{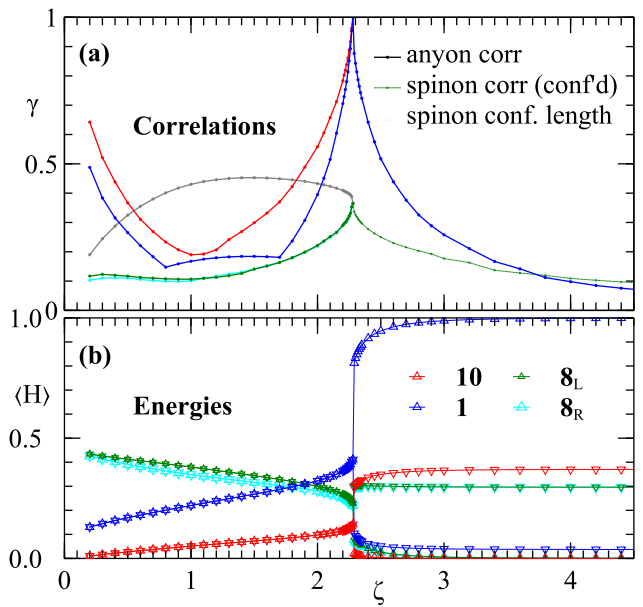}
\caption{\textbf{(a)} Correlation lengths for trivial and anyonic correlations vs.\ the
folded trimer weight $\zeta$ for the $\mathrm{SU}(3)$ model, where trimers
have been replaced by $\mathrm{SU}(3)$ singlets (i.e.\ fully antisymmetric
states), cf.\ also Fig.~\ref{fig:zeta-diag}.
The color coding used is the same as in Fig.~\ref{fig:interpol}. We again
identify a topological phase in the center, which transitions to a
critical phase for $\zeta\to0$, and to a symmetry-broken phase around
$\zeta_c\approx2.28$. In the symmetry broken phase on the right, spinons become
confined, and the associated spinon confinement length is shown by open
(green) circles. It first equals the two-point (trivial) correlation
length, and for large $\zeta\gtrsim3.7$ the (now unphysical) correlation
length between spinons. Data has been obtained with iMPS truncation
threshold $\eta=10^{-4}2^{-14}\approx6.10\times 10^{-9}$.
\textbf{(b)}  Variational energies vs.\ $\zeta$ for the different $\mathrm{SU}(3)$
invariant terms on up ($\vartriangle$) and down ($\triangledown$)
triangles, see text.
}
\label{fig:su3-joint}
\end{figure}

Second, we have studied the dependence of the phase diagram on the folded
trimer weight $\zeta$ at the $\mathrm{SU}(3)$ point $\mathcal P_3$.  The
results for the correlation lengths are shown in Fig.~\ref{fig:su3-joint}a
(we again plot $\gamma=e^{-1/\xi}$, with the same color coding for the different
anyon sectors as in Fig.~\ref{fig:interpol}): We again
identify a topological phase for $\zeta<\zeta_c\approx2.28$,
a symmetry broken valence bond crystal phase
for $\zeta>\zeta_c$, and a critical phase at or around $\zeta=0$, which
can be understood using the same qualitative picture as before for the
trimer model.  We find that the phase transitions out of the topological
phase are driven by diverging vison correlations (i.e.\ vison
condensation), just as for the trimer model, even though in the center of
the topological phase around $\zeta=1$, dyonic correlations (which include a
spinon contribution) are dominating. 
At the phase transition, the visons condense, leading to the confinement
of spinons in the symmetry broken crystalline phase.
Fig.~\ref{fig:su3-joint}a also shows the associated spinon confinement
length; interestingly, this length scale is initially dominated by the
equally divergent trivial correlations, while around $\zeta\gtrsim3.7$, the 
correlation length between spinons (which is in principle
unphysical, as spinons are now confined) becomes dominant.\footnote{Using
the Cauchy-Schwarz inequality, one can see that the spinon correlation
length should be \emph{upper bounded} by the spinon confinement
length, cf.\ Ref.~\onlinecite{duivenvoorden:anyon-condensation}.}
Finally, let us point out that while on the one hand, the critical point
$\zeta_c\approx2.28$ did not change significantly as compared to the
trimer model (where $\zeta_c\approx2.19$),\footnote{Note that a similar
robustness was observed when comparing the effect of vison doping in dimer
vs.\ RVB states~\cite{iqbal:rvb-perturb}.}
the implications on how the
inequivalent ways to resolve folded trimers affect the phase is rather
different: Since at the
$\mathrm{SU}(3)$ point, all folded trimers are replaced by
the \emph{same} $\mathrm{SU}(3)$ singlet, the weight $\zeta$ which appears
when resolving a folded trimer as three distinct trimers is $\zeta=3$
rather than $\zeta=\sqrt{3}$, and thus, the corresponding $\mathrm{SU}(3)$
model is in the symmetry broken crystalline rather than the topological
phase, as indeed reported in Ref.~\onlinecite{kurecic:su3_sl}.

Let us now discuss parent Hamiltonians in the context of the
$\mathrm{SU}(3)$ model. There are several aspects: On the one hand, we are
interested in the Hamiltonian for the $\mathrm{SU}(3)$ model itself, but
on the other hand, we will also discuss how the parent Hamiltonians for
the trimer model are affected if we equip the trimers with actual
$\mathrm{SU}(3)$ singlets (or some other sufficiently symmetric tripartite
entangled state), while keeping different trimer configurations
orthogonal.  More generally, we will discuss parent Hamiltonians along the
entire interpolation family $\mathcal P_1\to\mathcal P_2\to\mathcal P_3$,
where the points $\mathcal P_3$ and $\mathcal P_2$ ($\mathcal P_1$)
correspond to the two aforementioned cases.

Why is the Hamiltonian -- as discussed in Sec.~\ref{sec:z3-inj-ham} --
affected if we equip the trimers, for which we have
been hitherto using the arrow representation, with a tripartite entangled state? To
this end, consider the transition between the two configurations in
Fig.~\ref{fig:hamiltonian}b, and the corresponding Hamiltonian term: In the arrow
representaton (left), this is a one-hexagon move, where we just need to
change the arrows on the hexagon itself from blue to red.  On the other hand,
if we equip the trimers with entangled states, we must in addition update
the entangled states placed on top of the trimers. This can be done in
different ways, such as the one indicated by green arrows
(Fig.~\ref{fig:hamiltonian}b, right), but regardless of
how it is done, it requires to act on \emph{at least} one of the outer
vertices.  This situation occurs precisely if only the inner vertex of a
trimer lies on the hexagon, and the $\circ$ on the inner vertex is
changed in the transition: In that case, the 
degrees of freedom on the 
two outer vertices of the initial trimer, 
which are entangled in the initial state,
belong to two different trimers after the move, and are
thus part of two different entangled states: Hence, disentangling the degrees
of freeom 
requires to act on at least one of them. In all other cases, the trimer
pattern can be changed without touching the vertices outside the hexagon,
since the outside vertices remain part of the same trimer.\footnote{Note
that we assume that entangled state possesses a symmetry with respect to
exchanging the outer vertices of the trimer, cf.\
footnote~\ref{foot:state-on-trimers-sym}.} For
one-hexagon moves, in the worst case (Fig.~\ref{fig:hamiltonian}c) this requires to act
on $8$ spins, while the required two-hexagon move can be implemented by
acting on $12=11+1$ spins (two hexagons plus the vertex marked by the dashed
green circle in Fig.~\ref{fig:hamiltonian}a). Note that since the
local terms $h_{\varhexagon}$ and $h_{\doublehexagon}$ in the Hamiltonian
are of the form $h_\bullet=\openone-\sum\ket{\chi_k}\bra{\chi_k}$, with the
$\ket{\chi_k}$ the
superposition of coupled configurations (and thus $8$-or 12-local),
the total Hamiltonian can indeed be expressed as a sum of $8$-local or 
$12$-local terms, respectively.

Let us now return to the Hamiltonian along the interpolation $\mathcal
P_1\to\mathcal P_2\to\mathcal P_3$. Along the line $\mathcal P_1\to \mathcal
P_2$, the model corresponds to the trimer model, where the trimers have
been replaced by different entangled states depending on the interpolation
parameter.  Thus, our discussion above applies to the whole interpolation,
and implies the existence of an $8$-body and $12$-body Hamiltonian
acting around one and two hexagons, respectively, depending on whether we
are willing to accept an additional isolated ``ice'' sector on top of the
topologically degenerate ground
space.\footnote{\label{foot:state-on-trimers-sym}Note that the state which we
put on the trimers (footnote \ref{foot:trimer-state}) is antisymmetric under exchanging the outer vertices,
and thus, a Hamiltonian acting on one/two outer vertices as shown in
Fig.~\ref{fig:hamiltonian} is sufficient to update trimer configurations.}
As we move further and interpolate $\mathcal P_2\to \mathcal P_3$,
orthogonality of different trimer configurations is lost.  However, since
the arrow information in $\mathcal P_2$ is \emph{smoothly} removed through
a ``filtering'' map $\Lambda=(1-\theta)\openone + \theta\ket{+}\bra{+}$
on the arrow degree of freedom (with $\ket+$ the even weight
superposition)~\cite{schuch:rvb-kagome,kurecic:su3_sl}, the PEPS at any
point along the interpolation is related to the state $\mathcal P_2$ by a
local \emph{invertible} map, except for the final point $\mathcal P_3$
itself.  Since the Hamiltonian is frustration free, this implies that we
can change the local terms in the Hamiltonian correspondingly (using the
inverse map) in a smooth way without changing the structure of the
topological ground space which remains $9$-fold degenerate. However, this
only works if we apply this transformation to each of the terms
$h_{\varhexagon},h_{\doublehexagon}=\openone - \sum
\ket{\chi_k}\bra{\chi_k}$ as a whole, rather than the terms
$\ket{\chi_k}\bra{\chi_k}$ individually, and thus yields a $12$-local or
$19$-local Hamiltonian instead,\footnote{\label{foot:cont-ham-1}Concretely, for a
wavefunction $\ket\Psi$
with Hamiltonian $\sum h_i$, where $h_i\ge0$ and $h_i\ket\Psi=0$, 
an invertible deformation $\ket{\Psi'}=\Lambda^{\otimes N}\ket{\Psi}$ on the
wavefunction corresponds to a deformed parent Hamiltonian
$h_i'=\big((\Lambda^{-1})^\dagger\big)^{\otimes k} h_i (\Lambda^{-1})^{\otimes
k}$, with the tensor product $^{\otimes k}$ acting on the sites on which
$h_i$ acts. Note that $h_i'$ can be replaced by a projector $h_i''$ with
the same kernel while keeping smoothness in the deformation (as eigenspace
projectors of analytic maps are analytic as
well~\cite[Thm.~6.1]{kato:pert-of-linear-operators}).  
See Ref.~\onlinecite{schuch:rvb-kagome} for a detailed discussion.}
just as for the $\mathrm{SU}(3)$ model of
Ref.~\onlinecite{kurecic:su3_sl}.
Finally, for the $\mathrm{SU}(3)$-point $\mathcal P_3$, we can construct 
a Hamiltonian which yet again consists of $12$- or $19$-body terms by
taking the limit of the parent Hamiltonian along the interpolation;\footnote{
The argument is similar to the previous footnote~\ref{foot:cont-ham-1}:
Since the ground states $\ket{\chi_k}$ are polynomials in the deformation
parameter and thus analytic also around the final point, it follows that
the ground space changes analytically and thus the limit of the
projector-valued parent Hamiltonians $h_i''$ above exists.}
 however, we cannot rule out that this
Hamiltonian exhibits additional ground states.\footnote{Parent
Hamiltonians can also be constructed by using that the tensor network on
one star (12 sites), seen as a map from virtual to physical system, has
constant rank everywhere except at the point $\mathcal P_3$, and thus, an
invertible map on disjoint stars maps it to the orthogonal trimer model,
which however yields a Hamiltonian with significantly larger
locality~\cite{schuch:rvb-kagome}. A similar argument, using that the PEPS
map on a star is $G$-injective, was used for the
$\mathrm{SU}(3)$ model in Ref.~\onlinecite{kurecic:su3_sl}, where it gave a
$19$-body Hamiltonian; the reason why we
can apply a much more direct argument which maps the model to the trimer point
and thus yields simpler parent Hamiltonians, while not requiring numerical
checks of $G$-injectivity on large patches, lies in the fact that unlike in
Ref.~\onlinecite{kurecic:su3_sl}, no big
entangled clusters appear in the superposition \eqref{eq:trimerstate},
since the $000 \equiv \bar{\bm 3}\otimes \bar{\bm 3}\otimes \bar{\bm 3}$
configuration is missing.}

Finally, we have investigated the behavior of the $\mathrm{SU}(3)$ wavefunction
as a function of $\zeta$ as an ansatz for $\mathrm{SU}(3)$ Hamiltonians
with $3$-body interactions on triangles. Any such $3$-body term can be
decomposed as a sum of four projections $h_{\bm 1}$, $h_{\bm{10}}$,
$h_{\bm{8_L}}$, and $h_{\bm{8_R}}$ onto the corresponding irreps (here,
$\bm{8_L}$ and $\bm{8_R}$ denote the $\bm 8$ irreps with angular momentum
$\pm2\pi/3$, respectively); in particular, the Heisenberg-type Hamiltonian
of Ref.~\onlinecite{corboz:suN-heisenberg-simplex-solids} (the sum over 
permutations of nearest neighbors) corresponds to
$h_{\mathrm{Heis},\vartriangle}=3(h_{\bm{10}}-h_{\bm 1})$.  The expectation
values for the corresponding terms are shown in Fig.~\ref{fig:su3-joint}b,
where energies for up and down triangles are plotted with the
corresponding symbol.  One can clearly  identify the symmetry breaking
phase transition, and the fact that in the crystalline phase, the
up-triangles hold the folded trimers and are thus in a
singlet.\footnote{The symmetry breaking pattern is controlled by biasing
the initial state of the iMPS boundary.} Unfortunately, we found that for
all $\mathrm{SU}(3)$-invariant three-body interactions which we
considered, either the $\zeta=0$ or the $\zeta\to\infty$ point provide the
lowest variational energy, making it unlikely that the wavefunction
accurately captures the way in which the physics of $\mathrm{SU}(3)$
models with three-body interactions across triangles depends on the
interactions.

\section{Conclusions}

In this work, we have introduced and studied quantum trimer models and
resonating $\mathrm{SU}(3)$-singlet models on the kagome lattice. We
have devised an arrow representation for the trimer model which provides a
direct mapping to $\mathbb Z_3$ loop models.  While the loop pattern is 
missing a configuration, we showed that the full space of $\mathbb
Z_3$-configurations (that is, $\mathbb Z_3$-injectivity) is recovered
under blocking. This allowed us to combine analytical tensor network tools
with a microscopic analysis to devise simple parent Hamiltonians with
$6$-body or $11$-body interactions: While the $11$-body terms gives rise to
the correct $9$-fold ground space degeneracy, restricting to $6$-body
terms alone only gives rise to six additional isolated ``ice'' states with
frozen trimers which do not couple to any of the remaining configurations
and which we thus expect to be strongly suppressed.

We have subsequently studied the phase diagram of the model by combining
analytical arguments with numerical study. We have found that a key role
is played by the relative weight $\zeta$ of folded (on-triangle) vs.\
unfolded trimers. For sufficiently large $\zeta$, the system displays
a conventional symmetry-broken phase in which all $\mathrm{SU}(3)$
singlets are localized on one type of triangles. In the intermediate
regime $0<\zeta<\zeta_c\approx 2.19$, we identified a topological phase
with $\mathbb Z_3$ topological order. Finally, for $\zeta=0$, we found an
additional $\mathrm{U}(1)$ symmetry, indicative of critical behavior which
we confirmed numerically.

We have finally equipped the trimer model with $\mathrm{SU}(3)$ singlets
and shown how these give rise to only slighly enlarged $8$- or $12$-body
Hamiltonians. Using this as a starting point, we devised an interpolation
where we erase the trimer (arrow) information, leaving us with a pure
resonating $\mathrm{SU}(3)$-singlet model with the fundamental
representation per site.  We numerically studied this model for $\zeta=1$
along the interpolation from the trimer point, and found that it is a
$\mathbb Z_3$ topological spin liquid in the same phase as the trimer
model and the $\mathbb Z_3$ RG fixed point loop model. Investigating the dependence on $\zeta$ resulted in a phase diagram
simlar to the one above, with only a slight change in $\zeta_c\approx
2.28$; yet this implies that a model where we were to place
$\mathrm{SU}(3)$-singlets in three different ways on any triangle would be
trivial rather than topologically ordered, unlike for the quantum trimer
model.

Several open questions remain. For one thing, it would be interesting to
understand if there is a connection between the geometry of the lattice
and the range of phases it can host, similar to the bipartite vs.\
non-bipartite phenomenon for $\mathrm{SU}(2)$ RVBs.  Similarly, the
extension of the trimer construction to $\mathrm{SU}(N)$ and $N$-mers
holds a wide range of interesting questions, from the nature of their
quantum order or the role played by the lattice to the effect of the
growing number of inequivalent trimer weights. Finally, our findings suggest
that our ansatz, even though it exhibits a symmetry broken phase, does not
capture the physics of $\mathrm{SU}(3)$ Heisenberg models beyond mean
field, leaving the quest for a suitable $\mathrm{SU}(3)$ ansatz open.

\begin{acknowledgments}
We acknowledge helpful discussions with Hong-Hao Tu.  This work has
received support from the European Union's Horizon 2020 program through
the ERC-StG WASCOSYS (No.~636201), and from the DFG (German Research
Foundation) under Germany's Excellence Strategy (EXC2111-390814868).
\end{acknowledgments}

\onecolumngrid
\vspace*{0.6cm}
\begin{center}
\rule{10cm}{.4pt}\\[-9pt]
\rule{12cm}{.4pt}\\[-9pt]
\rule{10cm}{.4pt}
\end{center}
\vspace*{0.6cm}
\twocolumngrid

\appendix

\section{Two-hexagon parent Hamiltonian\label{sec:app:2hexham}}

\subsection{Setting and goal}

\begin{figure}
\includegraphics[width=6cm]{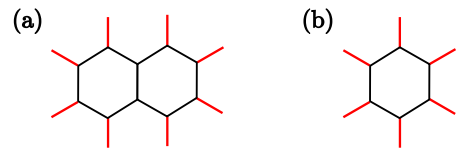}
\caption{\textbf{(a)} The two-hexagon parent Hamiltonian acts on the interior indices
of the two hexagons (black edges), and couples all interior configuations
compatible with the boundary configurations (red edges); note that the
latter can be uniquely inferred from the interior configurations. In
addition, Hamiltonian terms enforcing the vertex constraints are required.
\textbf{(b)} The one-hexagon Hamiltonian only acts on the interior degrees
of freedom of one hexagon. It leads to decoupled ``ice'' sectors
(Fig.~\ref{fig:hamiltonian}a), see Appendix~\ref{sec:app:1hexham}.
}
\label{fig:app:2hex-1hex}
\end{figure}

In this appendix, we show that the trimer model appears as the
topologically nine-fold degenerate ground state of a Hamiltonian 
\begin{equation}
H = \sum h_{\Yleft} + \sum h_{\doublehexagon}
\end{equation}
cf.~Eq.~\eqref{eq:ham-doublehex},
where $h_{\Yleft}$ and $h_{\doublehexagon}$ act on the degrees of freedom
adjacent to a vertex and to a pair of hexagons, respectively. As discussed
in the main text, $h_{\Yleft}$ ensures that the ground space is spanned by
valid $\mathbb Z_3$ loop configurations, while 
$h_{\doublehexagon}$ couples different loop configurations which only
differ on the two hexagons (that is, its ground space consists of properly
weighted superpositions of the corresponding configurations),
Fig.~\ref{fig:app:2hex-1hex}a.

What remains to be shown is that any two configurations in the same
topological sector can be coupled by a sequence of two-hexagon moves
$h_{\doublehexagon}$: By construction, all the topological sectors are
ground states of $\sum h_{\doublehexagon}$, and the fact that all
configurations are coupled implies that they can only appear in a
superposition of \emph{all} allowed loop patterns with the correct
relative weight, that is, there are no other ground states (which would
imply decoupled sectors of loop patterns).

As discussed in the main text, a parent Hamiltonian with terms $h'$ constructed
on a region sufficiently larger than the injective region of
Fig.~\ref{fig:injective}f has the correct ground space
structure~\cite{schuch:peps-sym,schuch:rvb-kagome,molnar:normal-peps-fundamentalthm}. On the
other hand, the ground space of this $h'$ is precisely spanned by the allowed
loop configurations on that region, where configurations which only
differ inside that region are coupled. Differently speaking, being able to
induce transitions between any two configurations on 
the region supporting $h'$ is sufficient to couple \emph{any} two configurations in the same
sector.

What remains to be shown is that two-hexagon moves allow to construct any
desired transition on a larger region. This is what we will do in this
appendix.

\subsection{Definitions and notation}

\begin{figure}
\centering
\includegraphics[width=6.5cm]{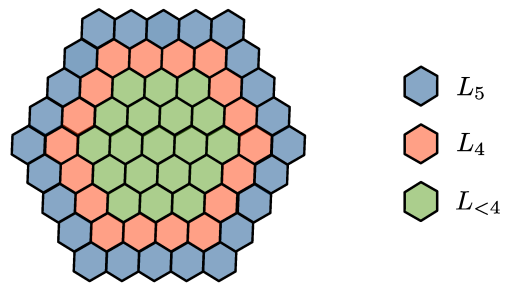}
\caption{Geometry considered in the proof of the two-hexagon parent
Hamiltonian. The blue and red hexagons form a $5$-ring and a $4$-ring,
respectively; a height configuration assigned to these hexagons is denoted
by $L_5$ and $L_4$. The green hexagons form a $3$-disc,
with configuration $L_{<4}=(L_3,L_2,L_1)$.
}
\label{fig:rings}
\end{figure}

For the purpose of the proof, we will consider hexagon-shaped blocks as
the one shown in Fig.~\ref{fig:rings}, and we will show that using
two-hexagon moves, any move on such hexagons (of any given size) can be
achieved. In particular, this region can be chosen large enough to contain
the aforementioned parent Hamiltonian $h'$, which yields all transitions
induced by $h'$, and thus implies the existence of transitions between all
configurations in the same topological sector.

Here, ``move'' denotes any transition between two loop configurations on
the edges inside the respective region,  in such a way
that both configurations are consistent as parts of a larger system. 
This is illustrated in Fig.~\ref{fig:app:2hex-1hex}a for a two-hexagon
move, which maps any configuration of the $11$ black edges to any other
configuration which is consistent with the same value of the surrounding red legs.
Note that the value of the red legs can be inferred from the
black degrees of freedom at the inside, i.e.\ such a move does not require
to access the exterior indices.

It will be convenient to work in the height representation of
configurations, that is, with plaquette variables.  Let us introduce some
language: We will denote a hexagon-shaped ring of $6n-6$ hexagons, such as
the blue and red rings in Fig.~\ref{fig:rings}, as an \emph{$n$-ring} ($n$
counts the hexagons at one edge) -- e.g., 
the blue hexagons in Fig.~\ref{fig:rings} forms a $5$-ring, and the red
hexagons form a $4$-ring.  The union of all $m$-rings with $m\le n$ will
be called an $n$-disc; e.g., the green region is a $3$-disc.  In the
height representation, each hexagon is assigned a height value, with the
constraint that three identical heights adjacent to each other are
forbidden, Fig.~\ref{fig:injective}b.  Then, a \emph{move} acting on a
region consists of replacing the height configuration inside that region
by another one, keeping the outside configuration fixed, in such a way
that no forbidden configuration appears.  In particular, a two-hexagon
move amounts to changing the height value of two adjacent plaquettes, and a
single-hexagon move that of a single hexagon.
Note that we only consider
concentric rings as in Fig.~\ref{fig:rings}, making the notion of
$n$-rings and $n$-discs unique.

In the following, we will show inducively that if two configurations on an
$n$-disc agree on the exterior $n$-ring, we can make them agree on their interior
$(n-1)$-discs by a sequence of local moves acting on two adjacent hexagons
each. Since the region on which the original parent Hamiltonian $h'$ acts
is contained in an $n$-disc for sufficiently large $n$ (and thus couples
\emph{at most} all configurations on that $n$-disc), this implies that a
two-hexagon parent Hamiltonian is sufficient to couple any two
valid height configurations.

Let us introduce a notation for height configurations on the concentric
$n$-rings and $n$-discs.  First, we denote the height configuration on the
$n$-ring  by $L_n$.  We say that two configurations 
$L_n$ and $L_{n-1}$ are \emph{consistent} if no forbidden 
configurations (three equal adjacent heights, Fig.~\ref{fig:injective}b)
appear. For two consistent configurations $L_n$ and $L_{n-1}$, we denote
the joint configuration on the $n$-ring and the $(n-1)$-ring by
$(L_{n},L_{n-1})$, and so forth.
Finally, $L_{<n}=(L_{n-1},L_{n-2},\dots,L_1)$ denotes the joint 
configuration of the $(n-1)$-disc; and thus  e.g.\ 
$(L_{n+1},L_{<n+1})$ or $(L_{n+1}, L_n, L_{<n})$ the joint configuration
on the $(n+1)$-disc. Whenever we use this notation, this implies
consistency of the configuration.
For example, a configuration in Fig.~\ref{fig:rings} would be denoted by
$(L_{n+1},L_{n}, L_{<n})$ with $n=4$, where $L_{n+1}$ is the configuration on
the blue, $L_n$ the configuration on the red, and $L_{<n}$ the
configuration on the green hexagons.

When two configurations $(L_n, L_{<n})$ and $(L_n, L'_{<n})$ are related
by \emph{local moves} on the $(n-1)$-disc we write  $(L_n, L_{<n})
\longleftrightarrow (L_n, L'_{<n})$.  Here, local moves are those which are
generated by two-hexagon moves on the $(n-1)$-disc: That is, either
elementary
two-hexagon moves, or moves for which we have already shown
that they can be constructed from two-hexagon moves.  Clearly, 
being related by local moves is transitive.

\subsection{Overview of the proof}

We will prove the following:  

\begin{theorem}
For any $n\ge3$, $L_n$, and $L_{<n},L_{<n}'$ consistent with $L_n$, the
transition from $L_{<n}$ to $L_{<n}'$ can be realized through
local (two-hexagon) moves:
\begin{equation}
\forall L_n,L_{<n},L_{<n}':\ (L_n,L_{<n}) \longleftrightarrow (L_n,L_{<n}')\ .
\label{eq:app:induction-Hn}
\end{equation}
The moves only act on the inside, i.e., the $(n-1)$-disc, and leave $L_n$
unchanged throughout.
\end{theorem}

\begin{proof}
The proof will proceed by induction, with induction hypothesis
\eqref{eq:app:induction-Hn}.  That \eqref{eq:app:induction-Hn} holds for
$n=3$ can be checked by brute force.  Notably, this is the only step where
two-hexagon moves are required; we will get back to this point in
Appendix~\ref{sec:app:1hexham}.

Let us now prove the induction step, that is, if
\eqref{eq:app:induction-Hn} holds for $n\ge3$, it also holds for $n'=n+1$.
To this end, given configurations
$(L_{n+1}, L_n, L_{<n})$ and $(L_{n+1}, L'_n, L'_{<n})$ which we want to
connect through local moves, we first construct a sequence 
$L_n=L^{(0)}_n , L^{(1)}_n,\dots,L^{(N)}_n=L'_n$
with the following properties:
\begin{itemize}
\item[(i)] $L^{(i)}_n$ and $L^{(i+1)}_n$ differ only on a single hexagon.
\item[(ii)] For all $i$, $(L_{n+1}, L^{(i)}_n)$ is consistent.
\end{itemize}
The construction of this sequence is given in                                  
Section~\ref{sec:app:section_n_ring}. 
Next, for $i=1,\dots,N-1$, we show in Section~\ref{sec:app:n-1_ring} that
since $L_n^{(i)}$ and $L_n^{(i+1)}$ only differ on a single hexagon, we
can construct an $L_{<n}^{(i)}$ such that 
\begin{itemize}
\item[(iii)] Both $(L^{(i)}_n, L^{(i)}_{<n})$ and $(L^{(i+1)}_n,
L^{(i)}_{<n})$ are consistent.
\end{itemize}
Then, we have that
\begin{align*}
(L_{n+1},L_n,L_{<n}) 
&\stackrel{*}{\longleftrightarrow}
	(L_{n+1},L^{(0)}_n, L^{(0)}_{<n})\\
& \stackrel{\varhexagon}{\longleftrightarrow}
	(L_{n+1},L^{(1)}_n, L^{(0)}_{<n}) \\
&\stackrel{*}{\longleftrightarrow}
	(L_{n+1},L^{(1)}_n, L^{(1)}_{<n}) \\
& \stackrel{\varhexagon}{\longleftrightarrow}
	(L_{n+1},L^{(2)}_n, L^{(1)}_{<n}) \\
&\stackrel{}{\longleftrightarrow}\quad \cdots\cdots\\
&\stackrel{\varhexagon}{\longleftrightarrow}
	(L_{n+1},L^{(N)}_n, L^{(N-1)}_{<n})\\
& \stackrel{*}{\longleftrightarrow}
	(L_{n+1},L'_n, L'_{<n})\ .
\end{align*}  
Here, we have used the induction hypothesis 
\eqref{eq:app:induction-Hn} in the step marked
with a star, and a single-hexagon move on the one hexagon on which $L_n^{(i)}$
and $L_n^{(i+1)}$ differ [condition (i)] in the steps marked with a hexagon.  Note that
conditions (ii) and (iii) imply that all intermediate configurations are
consistent.

This shows that if \eqref{eq:app:induction-Hn} holds for $n$, it also
holds for $n'=n+1$, and thus completes the proof.
\end{proof}

\subsection{Construction of $L^{(0)}_n,...,L^{(N)}_n$}
\label{sec:app:section_n_ring}

\begin{figure}
\includegraphics[width=.75\columnwidth]{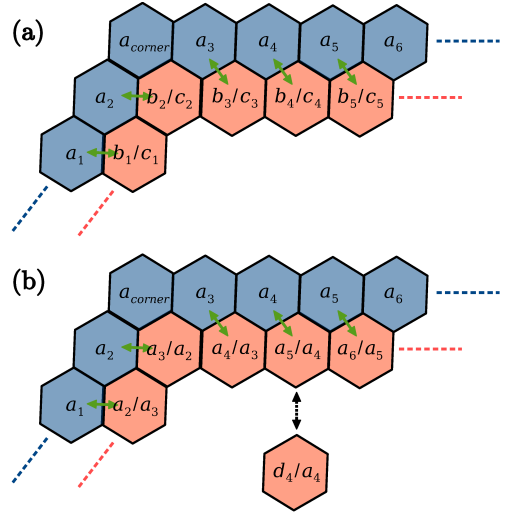}
\caption{
\textbf{(a)} Setting considered when constructing the sequence
$L_n^{(k)}$, see Section~\ref{sec:app:section_n_ring}. The configuration
$L_{n+1}$ of the $(n+1)$-ring (blue) is denoted by $a_i$, and those of the
two configurations $L_n$ and $L_n'$ of the $n$-ring (red) which we want to
connect by $b_i$ and $c_i$, respectively.  Green arrows indicate how the
plaquette labels $i$ are assigned relatively between the two rings.
\textbf{(b)} A setting $L_n$, $L_n'$ where all plaquettes are unflippable 
[where $b_{i-1}=c_i=a_i$ for all $i$, condition
\eqref{eq:app:unflippable-2}]. To resolve this situation, we insert an
additional configuration where we change one plaquette in the initial
state as indicated, with $d_4\ne a_4,a_5$. The new configuration now has
flippable plaquettes.}
\label{fig:n_rings}
\end{figure}

Here, we show the following: Given configurations $L_{n+1}$, $L_n$, and
$L_n'$, such that $L_{n+1}$ is consistent with both $L_n$ and $L_n'$, we
construct a sequence 
$L_n=L^{(0)}_n , L^{(1)}_n,\dots,L^{(N)}_n=L'_n$ such that all $L_n^{(k)}$
are consistent with $L_{n+1}$, and consecutive $L^{(k)}_n$ only differ on
a single hexagon.

Our construction will proceed sequentially, starting with $k=0$, and
$L^{(0)}_n=L_n$.  In each step -- labelled by $k$ -- we compare
$L^{(k)}_n$ with $L_n'$. We denote the height values of $L_{n+1}$
by $a_i$ and the height values of $L_{n}^{(k)}$ and $L_n'$ by $b_i$ and
$c_i$, respectively, as indicated in Fig.~\ref{fig:n_rings}a. 
Here, the plaquette labels $i$ are chosen such that the label of any
hexagon in the $n$-ring equals that of the hexagon in the $(n+1)$-ring to its
left, as seen from the center (cf.\ the green arrows in
Fig.~\ref{fig:n_rings}a).  The height values of the
corners of the $(n+1)$-ring will not be required.
We now proceed by identifying a \emph{flippable} plaquette $i$, that is, 
a plaquette for which $b_i\ne c_i$
and where we can change $b_i$ to $c_i$ in a way where the resulting
configuration is consistent, and then define $L_n^{(k+1)}$ to be the
resulting configuration on the $n$-ring  (that is,
$L^{(k+1)}_n$ equals $L_n^{(k)}$ and
thus $b_j$ on all plaquettes $j$ except for $i$, which is set to $c_i$). 

Clearly, this protocol will succeed in transforming $L^{(0)}_n$ to $L_n'$
if we can make sure that at every step $k$, there is at least one flippable
plaquette, since at most all the $6n-6$ plaquettes on the $n$-ring need to be
flipped.  Let us thus consider under which condition a plaquette is
\emph{unflippable}.  
\begin{subequations}
\label{eq:app:unflippable}
This can happen in two ways: Either, we already have
\begin{equation}
\label{eq:app:unflippable-1}
b_i=c_i\ ,
\end{equation}
or after changing $b_i$ to $c_i$, the configuration becomes
inconsistent across one of the two vertices,
\begin{align}
\label{eq:app:unflippable-2}
b_{i-1} &= a_i = c_i \mbox{\qquad or}\\
\label{eq:app:unflippable-3}
c_i & = a_{i+1} =b_{i+1} \ .
\end{align}
\end{subequations}
If (and only if) either of the equations \eqref{eq:app:unflippable} is
satisfied, then the plaquette $i$ is unflippable.

Let us now analyze the situation where all hexagons are unflippable. 
First, if $b_i=c_i$ for all hexagons, then $L_n^{(k)}=L_n'$ and we are
done.  
Thus, we can pick an $i$ for which $b_i \neq c_i$, i.e.\
\eqref{eq:app:unflippable-1} fails. Hence, either
\eqref{eq:app:unflippable-2} or \eqref{eq:app:unflippable-3} has to hold.
Assume w.l.o.g.\ that \eqref{eq:app:unflippable-2} holds. Since
$(L_{n+1},L'_n)$ is consistent, $\{c_{i-1},a_i,c_i\}$ cannot all be equal,
which together with \eqref{eq:app:unflippable-2} implies
$b_{i-1}=a_i=c_i\ne c_{i-1}$. Specifically, this means that
\begin{equation}
b_{i-1} \neq c_{i-1}\mbox{\quad and\quad} c_{i-1}\ne a_i\ ,
\end{equation}
and thus, neither \eqref{eq:app:unflippable-1} nor
\eqref{eq:app:unflippable-3} can hold for the hexagon $i-1$, such that
\eqref{eq:app:unflippable-2} must hold for the hexagon $i-1$ as well. By
continuing this way, we find that if all hexagons are unflippable and 
not all $b_i=c_i$, then the following two properties hold:
\begin{enumerate}
\item[(a)] Either \eqref{eq:app:unflippable-2} holds for all $i$, or \eqref{eq:app:unflippable-3}
holds for all~$i$.
\item[(b)] For all $i$, $b_i\ne c_i$.
\end{enumerate}
The case where \eqref{eq:app:unflippable-2} holds for all $i$ is
illustrated in Fig.~\ref{fig:n_rings}b. Note that e.g.\ the validity of 
\eqref{eq:app:unflippable-2} for all $i$ implies that 
\eqref{eq:app:unflippable-3} cannot hold anywhere, since otherwise
forbidden configurations would appear.

One important point is that point (b) above tells us that the situation
where all hexagons are unflippable can only appear in the very beginning,
$k=0$: After the first step, $k\ge1$, some of the hexagons in $L^{(k)}_n$
have already been flipped and therefore equal those in $L_n'$, $b_i=c_i$.

So what if when considering $L_n\equiv L_n^{(0)}$ and $L_n'$, all plaquettes
are unflippable, and thus point (a) above is satisfied? 
We will deal with
that by creating an additional configuration $L_n^{(1)}$ (obtained by
flipping one plaquette in $L_n^{(0)}$) such that comparing $L_n^{(1)}$ and
$L_n'$, there is a flippable hexagon.  W.l.o.g., we will assume that
\eqref{eq:app:unflippable-2} holds for all $i$, Fig.~\ref{fig:n_rings}b;
note that this implies that \eqref{eq:app:unflippable-3} holds nowhere.
Then, 
we pick any hexagon $i$ on the $n$-ring and define $L^{(1)}_n$ to be equal to
$L^{(0)}_n$ everywhere except on hexagon $i$, to which we assign the value
$d_i\ne a_i,a_{i+1}$.  $(L_{n+1},L^{(1)}_n)$ is consistent, as can be seen
from Fig.~\ref{fig:n_rings}b, where $i=4$, i.e.\ the $a_5$ in the marked
hexagon is now replaced by $d_4$ as indicated. 
By construction, $L^{(1)}_n$ differs from $L^{(0)}_n$ on a single site. 
Finally, when comparing $L^{(1)}_n$ and $L'$, the plaquette $i+1$ is
flippable, since condition \eqref{eq:app:unflippable-2} no longer holds by
choice of $d_i$, \eqref{eq:app:unflippable-3} did not hold to start
with, and $b_i\ne c_i$.

\subsection{Construction of the $L^{(i)}_{n-1}$}
\label{sec:app:n-1_ring} 

In this appendix we show that, given $L_n\equiv L_n^{(i)}$ and $L_n'\equiv
L_n^{(i+1)}$ which differ on one hexagon, $n\ge3$, we can
construct $L_{<n}\equiv L_{<n}^{(i)}$ such that both $(L_n,L_{<n})$ and
$(L_n',L_{<n})$ are consistent.

\begin{figure}
\includegraphics[width=0.88\columnwidth]{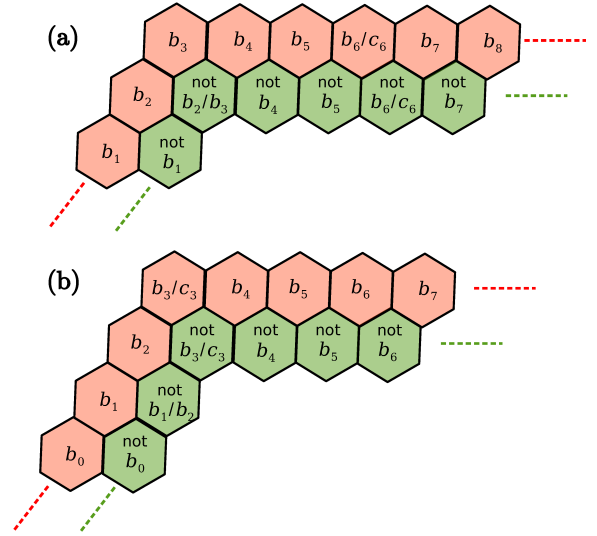}
\caption{
Construction of $L^{(i)}_{n-1}$ (green hexagons) in the case where
$L^{(i)}_n$ and $L^{(i+1)}_n$ (red hexagons) differ \textbf{(a)} at an
edge or \textbf{(b)} at a corner of the $n$-ring, see
Section~\ref{sec:app:n-1_ring}.
}
\label{fig:grow-rings}
\end{figure} 

First, consider the case where $n\ge4$. We set out by constructing
$L_{n-1}$ such that $(L_n,L_{n-1})$ and $(L_n',L_{n-1})$ are consistent.
There are two possibilities.  First,  the hexagon on which $L_n$ and
$L_n'$ differ is not at a corner of the $n$-ring, such as shown in
Fig.~\ref{fig:grow-rings}a for the outer (red) hexagon with values
$b_6/c_6$ (the values on which $L_n$ and $L_n'$ agree are denoted by
$b_i$). 
Then, construct $L_{n-1}$ as shown by the inner (green) hexagons in
Fig.~\ref{fig:grow-rings}a, where  "$\mathrm{not}\ x$" means that we can choose
any value except $x$, and "$\mathrm{not}\ x/y$" any value except $x$ and
$y$.  
By this choice, no three hexagons around a vertex can have the
same value for either $L_n$ or $L_n'$. 
In the second case, where the hexagon which differs is at a corner of
the $n$-ring, we construct $L_{n-1}$ as shown in
Fig.~\ref{fig:grow-rings}b. While now it is possible that the hexagon at
the corner of the $(n-1)$-ring has the value $b_2$, the choice of the
hexagon below as $\mathrm{not}\ b_1/b_2$ (rather than just $\mathrm{not}\
b_1$) ensures that also across that
vertex, the three hexagons cannot all have the same value.

Now that we have constructed $L_{n-1}$, we can use the same prescription to
construct $L_{n-2}$ such that $(L_{n-1},L_{n-2})$ is consistent (ignoring
the $L'$-part in the construction), and so forth.  Once we have arrived at
$L_3$, we can use the argument from the $G$-injectivity proof in the main
text, Fig.~\ref{fig:injective}e, to construct a consistent $L_{<3}$.  Putting
all the layers $(L_{n-1},L_{n-2},\dots,L_3,L_{<3})$ together, we obtain $L_{<n}$
such that both $(L_n,L_{<n})$ and $(L_n',L_{<n})$ are consistent.

The above argument fails for $n=3$ where we are given $L_3$ and $L_3'$ and
want to construct an $L_{2}$ consistent with both, since the $2$-ring
consists exclusively of corners. To cover the $n=3$ case, we will instead
provide a direct construction of $L_{<3}$, in close analogy to the
argument used in the main text to construct a consistent interior of an
arbitrary $3$-ring (Fig.~\ref{fig:injective}e).

\begin{figure}
\includegraphics[width=\columnwidth]{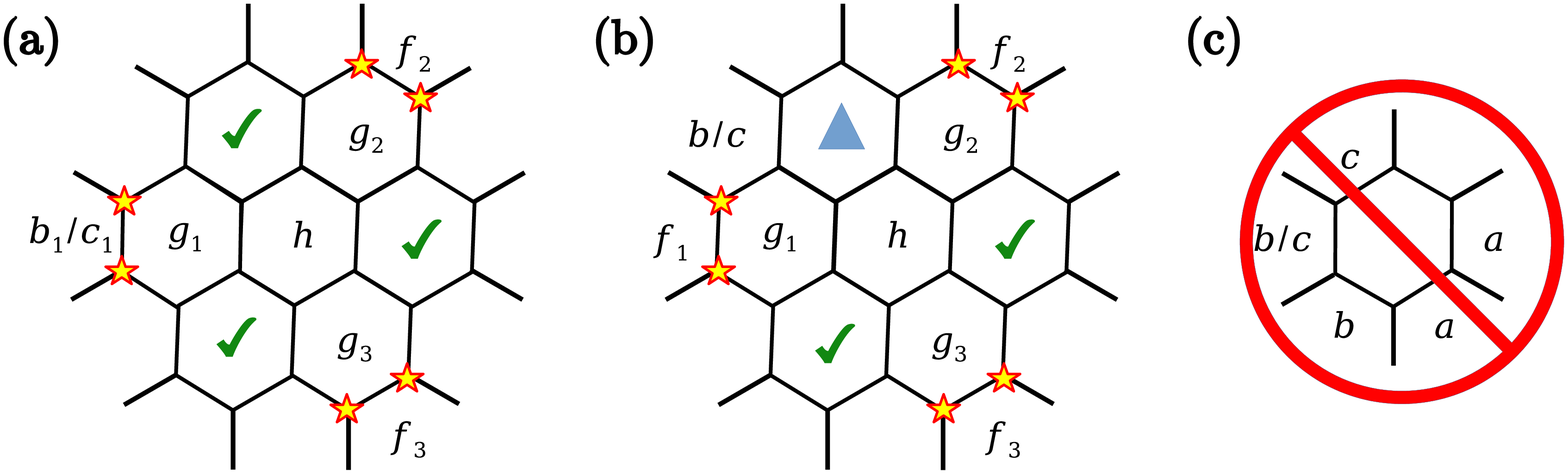}
\caption{
Construction of an interior configuration $L_{<3}$ consistent with two
outer configurations $L_3$ and $L_3'$ differing on one plaquette, see
Section~\ref{sec:app:n-1_ring}.  \textbf{(a)} Situation where the
differing plaquette is at the  corner. \textbf{(b)} Situation where the
differing plaquette is on the edge. \textbf{(c)} Forbidden configuration
(plus rotations/reflections) when the plaquette on the left can take two
values; here, $a,b,c$ are all different.
}
\label{fig:induction_basis}
\end{figure}

Yet again, we need to consider two cases. The first is where $L_3$ and
$L'_3$ differ at the corner of the $3$-ring, see
Fig.~\ref{fig:induction_basis}a.  We want to construct a configuration on
the interior that is consistent with both $b_1$ and $c_1$.
To this end, set $g_1 = \mathrm{not}(b_1,c_1)$,
$h=\mathrm{not}(g_1)$, 
$g_2 =\mathrm{not}(f_2,h)$ and  $g_3 =\mathrm{not}(f_3,h)$. Then
(\emph{i}) the vertices marked by stars cannot be in 
the forbidden configuration Fig.~\ref{fig:injective}b, i.e.\ three
identical adjacent values, and 
(\emph{ii}) the hexagons marked with a checkmark cannot be in the
forbidden configurations Fig.~\ref{fig:injective}d, and thus can be
assigned a consistent value.

If instead $L_3$ and $L_3'$ differ in the center of an edge, shown in
Fig.~\ref{fig:induction_basis}b, we need to give additional consideration
to the value of the hexagon marked with a triangle: Unlike for the hexagons
marked by a checkmark before, we now need to generalize
Fig.~\ref{fig:injective}d such as to rule out configurations which cannot be
completed in a way
consistent with both $b$ \emph{and} $c$ at the given plaquette. Those
configurations are of the form shown in Fig.~\ref{fig:induction_basis}c,
with $a,b,c$ all different, or variants thereof (reflection and/or
exchanging $b$ and $c$).  Now, in Fig.~\ref{fig:induction_basis}b, choose
$g_2 = \mathrm{not}(f_2, \mathrm{not}(b,c))$. This rules out the
configuration~\ref{fig:induction_basis}c around the hexagon marked by the
triangle, since there the hexagon opposite of $b/c$ has the value
$a=\mathrm{not}(b,c)$. Further, choose $h=\mathrm{not}(g_2)$, $g_1
=\mathrm{not}(f_1,h)$, and $g_3 =\mathrm{not}(f_3,h)$. Then, 
(\emph{i}) the vertices marked by stars cannot be in 
the forbidden configuration Fig.~\ref{fig:injective}b,  and
(\emph{ii}) we can fill
the hexagons marked with the triangle and the checkmarks in a consistent
way, as we have avoided the configurations Fig.~\ref{fig:induction_basis}c
and Fig.~\ref{fig:injective}d, respectively.  This completes the proof.

\section{Single hexagon Hamiltonian\label{sec:app:1hexham}}

\begin{figure}[t]
\includegraphics[width=8cm]{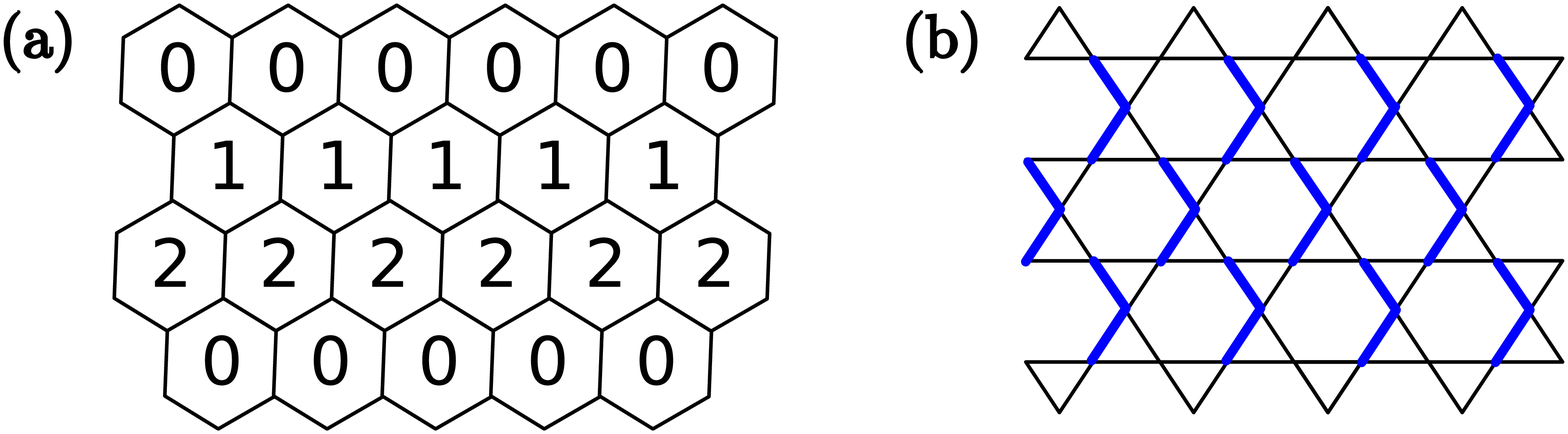}
\caption{Crystalline configuration \textbf{(a)} in the height
representation and \textbf{(b)} in the trimer representation. These
configurations cannot be changed by one-hexagon moves alone.}
\label{fig:staggered_states}
\end{figure}

In this appendix, we consider parent Hamiltonians which only consist of
one-hexagon moves, cf.~Fig.~\ref{fig:app:2hex-1hex}b (together with terms ensuring
the $\mathbb Z_3$ vertex constraints) and show that, with the exception of
the $6$ symmetry-related crystalline configurations in
Fig.~\ref{fig:staggered_states} (see also
Fig.~\ref{fig:hamiltonian}a), all other configurations are coupled
by one-hexagon moves.

To start with, it is straightforward to see that the stripe height
configuration in Fig.~\ref{fig:staggered_states}a, corresponding to the crystalline
trimer pattern in Fig.~\ref{fig:staggered_states}b, cannot be changed by one-hexagon
moves (see Fig.~\ref{fig:hamiltonian}a for a two-hexagon move melting such a
cluster), and is thus decoupled from any other configuration; we will
refer to the corresponding height configuration as ``crystalline'' as well. There are
altogether $6$ such configuration related by the lattice rotation symmetry.
However, as we will
see in the following, these are the \emph{only} such
configurations: All other configurations (within the same sector) can be
transformed into each other by one-hexagon moves.  First, recall 
from Appendix~\ref{sec:app:2hexham}
that the
only step where we required two-hexagon moves was the start of the
induction, that is, Eq.~\eqref{eq:app:induction-Hn} for $n=3$:
\begin{equation}
(L_3,L_{<3})\longleftrightarrow(L_3,L_{<3}')
 \mbox{\ for all\ } L_3, L_{<3}, L_{<3}'\ .
\label{eq:app:L3move}
\end{equation}
In the following, we will show that we can always implement
such a move \emph{anywhere} in the system as long as \emph{somewhere}
there is a place where the system is not in the crystalline configuration.
In particular, this implies that in order to only keep the $9$
topological ground states, it is sufficient if we include a \emph{single}
two-hexagon term (which can locally melt the crystal) \emph{somewhere} in the
system.

\begin{figure}
\includegraphics[width=8cm]{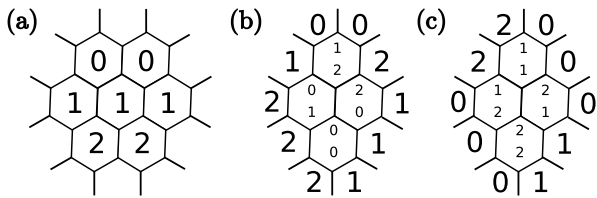}
\caption{\textbf{(a)} ``Frozen'' configurations which cannot be
changed by one-hexagon moves on the inner $2$-disc
for certain boundary conditions.
\textbf{(b,c)} 
The boundary configurations (up to symmetry) at the intersection of two
$2$-discs (Fig.~\ref{fig:shifting_hexagons}) which only allow for two
interior assignments (top/bottom); all other boundaries have at least
three interior assignments.}
\label{fig:computer_results}
\end{figure}

We are thus interested in the situation where we cannot perform a
transformation $(L_3,L_{<3})\onehexmove(L_3,L_{<3}')$ by a sequence of
single-hexagon moves (denoted $\onehexmove$) on the inner $2$-disc (i.e.,
$L_{<3}$ vs.\ $L_{<3}'$) alone. An exhaustive search shows that for any
given $L_3$, there are exactly two possibilities: Either \emph{all} interior
configurations $L_{<3}$, $L_{<3}'$ are connected,
$(L_3,L_{<3})\onehexmove(L_3,L_{<3}')$, or there is \emph{a single}
configuration $\hat L_{<3}$ which is cannot be connected to \emph{any}
other configuration $L_{<3}$, while \emph{all other} configurations
$L_{<3},L_{<3}'\ne \hat L_{<3}$ are still connected by one-hexagon moves,
$(L_3,L_{<3})\onehexmove(L_3,L_{<3}')$.  Moreover, such an $\hat L_{<3}$
is necessarily of the form shown in Fig.~\ref{fig:computer_results}a (up
to symmetry).  We will call such configuration -- which are the only ones
which require two-hexagon moves to change them -- \emph{frozen}
configurations.  Note that being frozen is a joint property of
$(L_3,\hat L_{<3})$ (as the allowed moves on the interior $2$-disc depend on $L_3$).  

Clearly, if all $2$-discs are frozen, the system must be in on of the
crystalline states, Fig.~\ref{fig:staggered_states}a. Thus, let us
consider the scenario where at least one $2$-disc is in a non-frozen
configuration.  We will now show that this allows us to ``melt'' frozen
configurations anywhere in the system with only one-hexagon moves.

\begin{figure}[b]
\includegraphics[width=8cm]{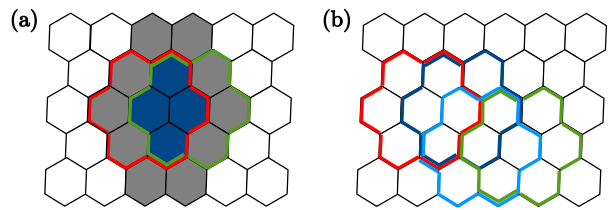}
\caption{Setup for ``melting'' frozen $2$-discs by one-hexagon moves (see
text). \textbf{(a)} The frozen left (red) $2$-disc can be melted if the
adjacent right (green) $2$-disc is not frozen, by updating the
intersection (blue) through one-hexagon moves on the right $2$-disc.
\textbf{(b)} If the frozen $2$-disc (left, red) is inside a frozen
cluster, it is connected through a path of frozen $2$-discs (blue) to a
non-frozen $2$-disc (right, green), and one applies (a) iteratively.
}
\label{fig:shifting_hexagons}
\end{figure}

To start with, consider the situation in
Fig.~\ref{fig:shifting_hexagons}a, where the left (red) $2$-disc is
frozen, and the right (green) $2$-disc isn't. Our goal is to change the
intersection of the two discs (blue) to a different configuration: As
there is only a single frozen configuration for a given boundary, this
would automatically ``melt'' the left $2$-disc and allow us to subsequently transform it
to any desired configuration (except the original frozen one) by one-hexagon moves alone.  Since the right
$2$-disc is not frozen, changing the intersection region is possible by
one-hexagon moves
on the right $2$-disc, \emph{unless} the desired configuration on the
intersection (blue) would amount to the frozen configuration of the right
$2$-disc. However, an exhaustive search reveals that -- given a fixed
boundary condition (gray) to the blue intersection -- the only cases (up
to symmetry) where there are only two choices for the blue intersection
are those in Fig.~\ref{fig:shifting_hexagons}bc, and neither of them is is
consistent with both a frozen left and right $2$-disc in the two
configurations.
For all other boundary conditions, there are at least three
choices for the blue intersection, so there must be one for which neither
of the two $2$-discs is frozen (as the frozen configurations $\hat L_{<3}$ are unique).
We can now use one-hexagon moves on the right $2$-disc to transform the
blue intersection into such a configuration (restoring the plaquettes
outside the intersection to their original state); then, the
left $2$-disc is no longer in the frozen configuration and can be
transformed to any other configuration by one-hexagon moves.

In case we want to change a frozen $2$-disc with no non-frozen $2$-disc
adjacent to it, we choose a path of overlapping frozen $2$-discs
connecting it to a non-frozen $2$-disc somewhere in the system; such a
path is illustrated in Fig.~\ref{fig:shifting_hexagons}b. We can then
start from the non-frozen $2$-disc and melt the adjacent $2$-disc as
described above, and continue our way though the chain of $2$-discs until
we arrive at the original $2$-disc which we want to melt.  Importantly,
while moving through the chain of $2$-discs, we can always leave the
system in its original state as we move on, as we explain using the
diagram in Fig.~\ref{fig:shifting_hexagons}a: (\emph{i}) In the first step of the
procedure, we are precisely in the situation as before, where the right
$2$-disc is not frozen and the part outside the intersection is restored
to its original state. (\emph{ii}) In the further steps, however, we start
in a situation where initially the left $2$-disc is frozen while the right
$2$-disc isn't any more, and at the end of the step, we want to leave the
outside part of the right $2$-disc in the residual part of the original
frozen configuration (though whether it is actually frozen might depend on
the environment). We can now use the same argument as before to see
that 
there is at least one configuration on the intersection
which is consistent with neither the left nor the right $2$-disc being
frozen; we can thus use moves on the right $2$-disc to transform the
outside part to the configuration it had in the original (frozen) configuration, and
the intersection to the corresponding third configuration. We thus see
that we can use this scheme to melt any frozen $2$-disc inside a frozen
cluster by acting along a one-dimensional path which connects it with the
outside of the cluster, while restoring all plaquettes outside the
melted $2$-disc to their original state.


\begin{thebibliography}{34}%
\makeatletter
\providecommand \@ifxundefined [1]{%
 \@ifx{#1\undefined}
}%
\providecommand \@ifnum [1]{%
 \ifnum #1\expandafter \@firstoftwo
 \else \expandafter \@secondoftwo
 \fi
}%
\providecommand \@ifx [1]{%
 \ifx #1\expandafter \@firstoftwo
 \else \expandafter \@secondoftwo
 \fi
}%
\providecommand \natexlab [1]{#1}%
\providecommand \enquote  [1]{``#1''}%
\providecommand \bibnamefont  [1]{#1}%
\providecommand \bibfnamefont [1]{#1}%
\providecommand \citenamefont [1]{#1}%
\providecommand \href@noop [0]{\@secondoftwo}%
\providecommand \href [0]{\begingroup \@sanitize@url \@href}%
\providecommand \@href[1]{\@@startlink{#1}\@@href}%
\providecommand \@@href[1]{\endgroup#1\@@endlink}%
\providecommand \@sanitize@url [0]{\catcode `\\12\catcode `\$12\catcode
  `\&12\catcode `\#12\catcode `\^12\catcode `\_12\catcode `\%12\relax}%
\providecommand \@@startlink[1]{}%
\providecommand \@@endlink[0]{}%
\providecommand \url  [0]{\begingroup\@sanitize@url \@url }%
\providecommand \@url [1]{\endgroup\@href {#1}{\urlprefix }}%
\providecommand \urlprefix  [0]{URL }%
\providecommand \Eprint [0]{\href }%
\providecommand \doibase [0]{https://doi.org/}%
\providecommand \selectlanguage [0]{\@gobble}%
\providecommand \bibinfo  [0]{\@secondoftwo}%
\providecommand \bibfield  [0]{\@secondoftwo}%
\providecommand \translation [1]{[#1]}%
\providecommand \BibitemOpen [0]{}%
\providecommand \bibitemStop [0]{}%
\providecommand \bibitemNoStop [0]{.\EOS\space}%
\providecommand \EOS [0]{\spacefactor3000\relax}%
\providecommand \BibitemShut  [1]{\csname bibitem#1\endcsname}%
\let\auto@bib@innerbib\@empty
\bibitem [{\citenamefont {Balents}(2010)}]{balents:spin-liquid-review-2010}%
  \BibitemOpen
  \bibfield  {author} {\bibinfo {author} {\bibfnamefont {L.}~\bibnamefont
  {Balents}},\ }\bibfield  {title} {\bibinfo {title} {\emph {Spin liquids in
  frustrated magnets}},\ }\href {https://doi.org/10.1038/nature08917}
  {\bibfield  {journal} {\bibinfo  {journal} {Nature}\ }\textbf {\bibinfo
  {volume} {464}},\ \bibinfo {pages} {199} (\bibinfo {year}
  {2010})}\BibitemShut {NoStop}%
\bibitem [{\citenamefont {Savary}\ and\ \citenamefont
  {Balents}(2017)}]{savary:spin-liquids}%
  \BibitemOpen
  \bibfield  {author} {\bibinfo {author} {\bibfnamefont {L.}~\bibnamefont
  {Savary}}\ and\ \bibinfo {author} {\bibfnamefont {L.}~\bibnamefont
  {Balents}},\ }\bibfield  {title} {\bibinfo {title} {\emph {Quantum Spin
  Liquids}},\ }\href {https://doi.org/10.1088/0034-4885/80/1/016502} {\bibfield
   {journal} {\bibinfo  {journal} {Rep. Prog. Phys.}\ }\textbf {\bibinfo
  {volume} {80}},\ \bibinfo {pages} {016502} (\bibinfo {year} {2017})},\
  \Eprint {https://arxiv.org/abs/arXiv:1601.03742} {arXiv:1601.03742}
  \BibitemShut {NoStop}%
\bibitem [{\citenamefont {Knolle}\ and\ \citenamefont
  {Moessner}(2019)}]{knolle:spin-liquids}%
  \BibitemOpen
  \bibfield  {author} {\bibinfo {author} {\bibfnamefont {J.}~\bibnamefont
  {Knolle}}\ and\ \bibinfo {author} {\bibfnamefont {R.}~\bibnamefont
  {Moessner}},\ }\bibfield  {title} {\bibinfo {title} {\emph {A Field Guide to
  Spin Liquids}},\ }\href
  {https://doi.org/10.1146/annurev-conmatphys-031218-013401} {\bibfield
  {journal} {\bibinfo  {journal} {Annu. Rev. Condens. Matter Phys.}\ }\textbf
  {\bibinfo {volume} {10}},\ \bibinfo {pages} {451} (\bibinfo {year} {2019})},\
  \Eprint {https://arxiv.org/abs/arXiv:1804.02037} {arXiv:1804.02037}
  \BibitemShut {NoStop}%
\bibitem [{\citenamefont {Anderson}(1973)}]{anderson:rvb}%
  \BibitemOpen
  \bibfield  {author} {\bibinfo {author} {\bibfnamefont {P.~W.}\ \bibnamefont
  {Anderson}},\ }\bibfield  {title} {\bibinfo {title} {\emph {Resonating
  valence bonds: A new kind of insulator?}},\ }\href@noop {} {\bibfield
  {journal} {\bibinfo  {journal} {Mater. Res. Bull.}\ }\textbf {\bibinfo
  {volume} {8}},\ \bibinfo {pages} {153} (\bibinfo {year} {1973})}\BibitemShut
  {NoStop}%
\bibitem [{\citenamefont {Moessner}\ and\ \citenamefont
  {Raman}(2008)}]{moessner:quantum-dimer-models}%
  \BibitemOpen
  \bibfield  {author} {\bibinfo {author} {\bibfnamefont {R.}~\bibnamefont
  {Moessner}}\ and\ \bibinfo {author} {\bibfnamefont {K.~S.}\ \bibnamefont
  {Raman}},\ }\bibfield  {title} {\bibinfo {title} {\emph {Quantum dimer
  models}},\ }\href@noop {} {\  (\bibinfo {year} {2008})},\ \Eprint
  {https://arxiv.org/abs/arXiv:0809.3051} {arXiv:0809.3051} \BibitemShut
  {NoStop}%
\bibitem [{\citenamefont {Fradkin}(2013)}]{fradkin2013field}%
  \BibitemOpen
  \bibfield  {author} {\bibinfo {author} {\bibfnamefont {E.}~\bibnamefont
  {Fradkin}},\ }\href {https://books.google.de/books?id=x7\_6MX4ye\_wC} {\emph
  {\bibinfo {title} {Field Theories of Condensed Matter Physics}}},\ Field
  Theories of Condensed Matter Physics\ (\bibinfo  {publisher} {Cambridge
  University Press},\ \bibinfo {year} {2013})\BibitemShut {NoStop}%
\bibitem [{\citenamefont {Moessner}\ and\ \citenamefont
  {Sondhi}(2001)}]{moessner:dimer-triangular}%
  \BibitemOpen
  \bibfield  {author} {\bibinfo {author} {\bibfnamefont {R.}~\bibnamefont
  {Moessner}}\ and\ \bibinfo {author} {\bibfnamefont {S.~L.}\ \bibnamefont
  {Sondhi}},\ }\bibfield  {title} {\bibinfo {title} {\emph {An RVB phase in the
  triangular lattice quantum dimer model}},\ }\href@noop {} {\bibfield
  {journal} {\bibinfo  {journal} {Phys. Rev. Lett.}\ }\textbf {\bibinfo
  {volume} {86}},\ \bibinfo {pages} {1881} (\bibinfo {year} {2001})},\ \Eprint
  {https://arxiv.org/abs/cond-mat/0007378} {cond-mat/0007378} \BibitemShut
  {NoStop}%
\bibitem [{\citenamefont {Misguich}\ \emph {et~al.}(2002)\citenamefont
  {Misguich}, \citenamefont {Serban},\ and\ \citenamefont
  {Pasquier}}]{misguich:dimer-kagome}%
  \BibitemOpen
  \bibfield  {author} {\bibinfo {author} {\bibfnamefont {G.}~\bibnamefont
  {Misguich}}, \bibinfo {author} {\bibfnamefont {D.}~\bibnamefont {Serban}},\
  and\ \bibinfo {author} {\bibfnamefont {V.}~\bibnamefont {Pasquier}},\
  }\bibfield  {title} {\bibinfo {title} {\emph {Quantum Dimer Model on the
  Kagome Lattice: Solvable Dimer-Liquid and Ising Gauge Theory}},\ }\href@noop
  {} {\bibfield  {journal} {\bibinfo  {journal} {Phys. Rev. Lett.}\ }\textbf
  {\bibinfo {volume} {89}},\ \bibinfo {pages} {137202} (\bibinfo {year}
  {2002})},\ \Eprint {https://arxiv.org/abs/cond-mat/0204428}
  {cond-mat/0204428} \BibitemShut {NoStop}%
\bibitem [{\citenamefont {Elser}\ and\ \citenamefont
  {Zeng}(1993)}]{elser:rvb-arrow-representation}%
  \BibitemOpen
  \bibfield  {author} {\bibinfo {author} {\bibfnamefont {V.}~\bibnamefont
  {Elser}}\ and\ \bibinfo {author} {\bibfnamefont {C.}~\bibnamefont {Zeng}},\
  }\bibfield  {title} {\bibinfo {title} {\emph {kagome spin-1/2
  antiferromagnets in the hyperbolic plane}},\ }\href
  {https://doi.org/10.1103/PhysRevB.48.13647} {\bibfield  {journal} {\bibinfo
  {journal} {Phys. Rev. B}\ }\textbf {\bibinfo {volume} {48}},\ \bibinfo
  {pages} {13647} (\bibinfo {year} {1993})}\BibitemShut {NoStop}%
\bibitem [{\citenamefont {Kitaev}(2003)}]{kitaev:toriccode}%
  \BibitemOpen
  \bibfield  {author} {\bibinfo {author} {\bibfnamefont {A.}~\bibnamefont
  {Kitaev}},\ }\bibfield  {title} {\bibinfo {title} {\emph {Fault-tolerant
  quantum computation by anyons}},\ }\href@noop {} {\bibfield  {journal}
  {\bibinfo  {journal} {Ann. Phys.}\ }\textbf {\bibinfo {volume} {303}},\
  \bibinfo {pages} {2} (\bibinfo {year} {2003})},\ \Eprint
  {https://arxiv.org/abs/quant-ph/9707021} {quant-ph/9707021} \BibitemShut
  {NoStop}%
\bibitem [{\citenamefont {Norman}(2016)}]{norman:spin-liquid-review}%
  \BibitemOpen
  \bibfield  {author} {\bibinfo {author} {\bibfnamefont {M.~R.}\ \bibnamefont
  {Norman}},\ }\bibfield  {title} {\bibinfo {title} {\emph {Herbertsmithite and
  the Search for the Quantum Spin Liquid}},\ }\href
  {https://doi.org/10.1103/RevModPhys.88.041002} {\bibfield  {journal}
  {\bibinfo  {journal} {Rev. Mod. Phys.}\ }\textbf {\bibinfo {volume} {88}},\
  \bibinfo {pages} {041002} (\bibinfo {year} {2016})},\ \Eprint
  {https://arxiv.org/abs/arXiv:1604.03048} {arXiv:1604.03048} \BibitemShut
  {NoStop}%
\bibitem [{\citenamefont {Schuch}\ \emph {et~al.}(2012)\citenamefont {Schuch},
  \citenamefont {Poilblanc}, \citenamefont {Cirac},\ and\ \citenamefont
  {P{\'e}rez-Garc{\'\i}a}}]{schuch:rvb-kagome}%
  \BibitemOpen
  \bibfield  {author} {\bibinfo {author} {\bibfnamefont {N.}~\bibnamefont
  {Schuch}}, \bibinfo {author} {\bibfnamefont {D.}~\bibnamefont {Poilblanc}},
  \bibinfo {author} {\bibfnamefont {J.~I.}\ \bibnamefont {Cirac}},\ and\
  \bibinfo {author} {\bibfnamefont {D.}~\bibnamefont {P{\'e}rez-Garc{\'\i}a}},\
  }\bibfield  {title} {\bibinfo {title} {\emph {Resonating valence bond states
  in the PEPS formalism}},\ }\href@noop {} {\bibfield  {journal} {\bibinfo
  {journal} {Phys. Rev. B}\ }\textbf {\bibinfo {volume} {86}},\ \bibinfo
  {pages} {115108} (\bibinfo {year} {2012})},\ \Eprint
  {https://arxiv.org/abs/arXiv:1203.4816} {arXiv:1203.4816} \BibitemShut
  {NoStop}%
\bibitem [{\citenamefont {Zhou}\ \emph {et~al.}(2014)\citenamefont {Zhou},
  \citenamefont {Wildeboer},\ and\ \citenamefont
  {Seidel}}]{zhou:rvb-parent-onestar}%
  \BibitemOpen
  \bibfield  {author} {\bibinfo {author} {\bibfnamefont {Z.}~\bibnamefont
  {Zhou}}, \bibinfo {author} {\bibfnamefont {J.}~\bibnamefont {Wildeboer}},\
  and\ \bibinfo {author} {\bibfnamefont {A.}~\bibnamefont {Seidel}},\
  }\bibfield  {title} {\bibinfo {title} {\emph {Ground state uniqueness of the
  twelve site RVB spin-liquid parent Hamiltonian on the kagome lattice}},\
  }\href {https://doi.org/10.1103/PhysRevB.89.035123} {\bibfield  {journal}
  {\bibinfo  {journal} {Phys. Rev. B}\ }\textbf {\bibinfo {volume} {89}},\
  \bibinfo {pages} {035123} (\bibinfo {year} {2014})},\ \Eprint
  {https://arxiv.org/abs/arXiv:1310.8000} {arXiv:1310.8000} \BibitemShut
  {NoStop}%
\bibitem [{\citenamefont {Gorshkov}\ \emph {et~al.}(2010)\citenamefont
  {Gorshkov}, \citenamefont {Hermele}, \citenamefont {Gurarie}, \citenamefont
  {Xu}, \citenamefont {Julienne}, \citenamefont {Ye}, \citenamefont {Zoller},
  \citenamefont {Demler}, \citenamefont {Lukin},\ and\ \citenamefont
  {Rey}}]{gorshkov:suN-optical-lattice}%
  \BibitemOpen
  \bibfield  {author} {\bibinfo {author} {\bibfnamefont {A.~V.}\ \bibnamefont
  {Gorshkov}}, \bibinfo {author} {\bibfnamefont {M.}~\bibnamefont {Hermele}},
  \bibinfo {author} {\bibfnamefont {V.}~\bibnamefont {Gurarie}}, \bibinfo
  {author} {\bibfnamefont {C.}~\bibnamefont {Xu}}, \bibinfo {author}
  {\bibfnamefont {P.~S.}\ \bibnamefont {Julienne}}, \bibinfo {author}
  {\bibfnamefont {J.}~\bibnamefont {Ye}}, \bibinfo {author} {\bibfnamefont
  {P.}~\bibnamefont {Zoller}}, \bibinfo {author} {\bibfnamefont
  {E.}~\bibnamefont {Demler}}, \bibinfo {author} {\bibfnamefont {M.~D.}\
  \bibnamefont {Lukin}},\ and\ \bibinfo {author} {\bibfnamefont {A.~M.}\
  \bibnamefont {Rey}},\ }\bibfield  {title} {\bibinfo {title} {\emph
  {Two-orbital SU(N) magnetism with ultracold alkaline-earth atoms}},\ }\href
  {https://doi.org/10.1038/NPHYS1535} {\bibfield  {journal} {\bibinfo
  {journal} {Nature Phys.}\ }\textbf {\bibinfo {volume} {6}},\ \bibinfo {pages}
  {289} (\bibinfo {year} {2010})},\ \Eprint
  {https://arxiv.org/abs/arXiv:0905.2610} {arXiv:0905.2610} \BibitemShut
  {NoStop}%
\bibitem [{\citenamefont {Scazza}\ \emph {et~al.}(2014)\citenamefont {Scazza},
  \citenamefont {Hofrichter}, \citenamefont {H\"ofer}, \citenamefont {Groot},
  \citenamefont {Bloch},\ and\ \citenamefont
  {F\"olling}}]{scazza:suN-optical-lattice}%
  \BibitemOpen
  \bibfield  {author} {\bibinfo {author} {\bibfnamefont {F.}~\bibnamefont
  {Scazza}}, \bibinfo {author} {\bibfnamefont {C.}~\bibnamefont {Hofrichter}},
  \bibinfo {author} {\bibfnamefont {M.}~\bibnamefont {H\"ofer}}, \bibinfo
  {author} {\bibfnamefont {P.~C.~D.}\ \bibnamefont {Groot}}, \bibinfo {author}
  {\bibfnamefont {I.}~\bibnamefont {Bloch}},\ and\ \bibinfo {author}
  {\bibfnamefont {S.}~\bibnamefont {F\"olling}},\ }\bibfield  {title} {\bibinfo
  {title} {\emph {Observation of two-orbital spin-exchange interactions with
  ultracold SU(N)-symmetric fermions}},\ }\href
  {https://doi.org/10.1038/nphys3061} {\bibfield  {journal} {\bibinfo
  {journal} {Nature Phys.}\ }\textbf {\bibinfo {volume} {10}},\ \bibinfo
  {pages} {779} (\bibinfo {year} {2014})},\ \Eprint
  {https://arxiv.org/abs/arXiv:1403.4761} {arXiv:1403.4761} \BibitemShut
  {NoStop}%
\bibitem [{\citenamefont {Zhang}\ \emph {et~al.}(2014)\citenamefont {Zhang},
  \citenamefont {Bishof}, \citenamefont {Bromley}, \citenamefont {Kraus},
  \citenamefont {Safronova}, \citenamefont {Zoller}, \citenamefont {Rey},\ and\
  \citenamefont {Ye}}]{zhang:suN-experiment}%
  \BibitemOpen
  \bibfield  {author} {\bibinfo {author} {\bibfnamefont {X.}~\bibnamefont
  {Zhang}}, \bibinfo {author} {\bibfnamefont {M.}~\bibnamefont {Bishof}},
  \bibinfo {author} {\bibfnamefont {S.~L.}\ \bibnamefont {Bromley}}, \bibinfo
  {author} {\bibfnamefont {C.~V.}\ \bibnamefont {Kraus}}, \bibinfo {author}
  {\bibfnamefont {M.~S.}\ \bibnamefont {Safronova}}, \bibinfo {author}
  {\bibfnamefont {P.}~\bibnamefont {Zoller}}, \bibinfo {author} {\bibfnamefont
  {A.~M.}\ \bibnamefont {Rey}},\ and\ \bibinfo {author} {\bibfnamefont
  {J.}~\bibnamefont {Ye}},\ }\bibfield  {title} {\bibinfo {title} {\emph
  {Spectroscopic observation of SU(N)-symmetric interactions in Sr orbital
  magnetism}},\ }\href {https://doi.org/10.1126/science.1254978} {\bibfield
  {journal} {\bibinfo  {journal} {Science}\ }\textbf {\bibinfo {volume}
  {345}},\ \bibinfo {pages} {1467} (\bibinfo {year} {2014})},\ \Eprint
  {https://arxiv.org/abs/arXiv:1403.2964} {arXiv:1403.2964} \BibitemShut
  {NoStop}%
\bibitem [{\citenamefont {Lee}\ \emph {et~al.}(2017)\citenamefont {Lee},
  \citenamefont {Oh}, \citenamefont {Han},\ and\ \citenamefont
  {Katsura}}]{lee:resonating-trimer-state}%
  \BibitemOpen
  \bibfield  {author} {\bibinfo {author} {\bibfnamefont {H.}~\bibnamefont
  {Lee}}, \bibinfo {author} {\bibfnamefont {Y.}~\bibnamefont {Oh}}, \bibinfo
  {author} {\bibfnamefont {J.~H.}\ \bibnamefont {Han}},\ and\ \bibinfo {author}
  {\bibfnamefont {H.}~\bibnamefont {Katsura}},\ }\bibfield  {title} {\bibinfo
  {title} {\emph {Resonating Valence Bond States with Trimer Motifs}},\ }\href
  {https://doi.org/10.1103/PhysRevB.95.060413} {\bibfield  {journal} {\bibinfo
  {journal} {Phys. Rev. B}\ }\textbf {\bibinfo {volume} {95}},\ \bibinfo
  {pages} {060413} (\bibinfo {year} {2017})},\ \Eprint
  {https://arxiv.org/abs/arXiv:1612.06899} {arXiv:1612.06899} \BibitemShut
  {NoStop}%
\bibitem [{\citenamefont {Dong}\ \emph {et~al.}(2018)\citenamefont {Dong},
  \citenamefont {Chen},\ and\ \citenamefont
  {Tu}}]{dong:su3-trimer-squarelattice}%
  \BibitemOpen
  \bibfield  {author} {\bibinfo {author} {\bibfnamefont {X.-Y.}\ \bibnamefont
  {Dong}}, \bibinfo {author} {\bibfnamefont {J.-Y.}\ \bibnamefont {Chen}},\
  and\ \bibinfo {author} {\bibfnamefont {H.-H.}\ \bibnamefont {Tu}},\
  }\bibfield  {title} {\bibinfo {title} {\emph {SU(3) trimer
  resonating-valence-bond state on the square lattice}},\ }\href
  {https://doi.org/10.1103/PhysRevB.98.205117} {\bibfield  {journal} {\bibinfo
  {journal} {Phys. Rev. B}\ }\textbf {\bibinfo {volume} {98}},\ \bibinfo
  {pages} {205117} (\bibinfo {year} {2018})},\ \Eprint
  {https://arxiv.org/abs/1807.03254} {1807.03254} \BibitemShut {NoStop}%
\bibitem [{\citenamefont {Kurecic}\ \emph {et~al.}(2019)\citenamefont
  {Kurecic}, \citenamefont {Vanderstraeten},\ and\ \citenamefont
  {Schuch}}]{kurecic:su3_sl}%
  \BibitemOpen
  \bibfield  {author} {\bibinfo {author} {\bibfnamefont {I.}~\bibnamefont
  {Kurecic}}, \bibinfo {author} {\bibfnamefont {L.}~\bibnamefont
  {Vanderstraeten}},\ and\ \bibinfo {author} {\bibfnamefont {N.}~\bibnamefont
  {Schuch}},\ }\bibfield  {title} {\bibinfo {title} {\emph {A gapped SU(3) spin
  liquid with Z3 topological order}},\ }\href
  {https://doi.org/10.1103/PhysRevB.99.045116} {\bibfield  {journal} {\bibinfo
  {journal} {Phys. Rev. B}\ }\textbf {\bibinfo {volume} {99}},\ \bibinfo
  {pages} {045116} (\bibinfo {year} {2019})},\ \Eprint
  {https://arxiv.org/abs/arXiv:1805.11628} {arXiv:1805.11628} \BibitemShut
  {NoStop}%
\bibitem [{\citenamefont {{Schuch}}\ \emph {et~al.}(2010)\citenamefont
  {{Schuch}}, \citenamefont {{Cirac}},\ and\ \citenamefont
  {{P{\'e}rez-Garc{\'{\i}}a}}}]{schuch:peps-sym}%
  \BibitemOpen
  \bibfield  {author} {\bibinfo {author} {\bibfnamefont {N.}~\bibnamefont
  {{Schuch}}}, \bibinfo {author} {\bibfnamefont {I.}~\bibnamefont {{Cirac}}},\
  and\ \bibinfo {author} {\bibfnamefont {D.}~\bibnamefont
  {{P{\'e}rez-Garc{\'{\i}}a}}},\ }\bibfield  {title} {\bibinfo {title} {\emph
  {{PEPS as ground states: Degeneracy and topology}}},\ }\href
  {https://doi.org/10.1016/j.aop.2010.05.008} {\bibfield  {journal} {\bibinfo
  {journal} {Ann. Phys.}\ }\textbf {\bibinfo {volume} {325}},\ \bibinfo {pages}
  {2153} (\bibinfo {year} {2010})},\ \Eprint
  {https://arxiv.org/abs/arXiv:1001.3807} {arXiv:1001.3807} \BibitemShut
  {NoStop}%
\bibitem [{\citenamefont {Verstraete}\ and\ \citenamefont
  {Cirac}(2004{\natexlab{a}})}]{verstraete:mbc-peps}%
  \BibitemOpen
  \bibfield  {author} {\bibinfo {author} {\bibfnamefont {F.}~\bibnamefont
  {Verstraete}}\ and\ \bibinfo {author} {\bibfnamefont {J.~I.}\ \bibnamefont
  {Cirac}},\ }\bibfield  {title} {\bibinfo {title} {\emph {Valence Bond Solids
  for Quantum Computation}},\ }\href@noop {} {\bibfield  {journal} {\bibinfo
  {journal} {Phys.~Rev.~A}\ }\textbf {\bibinfo {volume} {70}},\ \bibinfo
  {pages} {060302} (\bibinfo {year} {2004}{\natexlab{a}})},\ \Eprint
  {https://arxiv.org/abs/quant-ph/0311130} {quant-ph/0311130} \BibitemShut
  {NoStop}%
\bibitem [{\citenamefont {Verstraete}\ and\ \citenamefont
  {Cirac}(2004{\natexlab{b}})}]{verstraete:2D-dmrg}%
  \BibitemOpen
  \bibfield  {author} {\bibinfo {author} {\bibfnamefont {F.}~\bibnamefont
  {Verstraete}}\ and\ \bibinfo {author} {\bibfnamefont {J.~I.}\ \bibnamefont
  {Cirac}},\ }\bibfield  {title} {\bibinfo {title} {\emph {Renormalization
  algorithms for Quantum-Many Body Systems in two and higher dimensions}},\
  }\href@noop {} {\  (\bibinfo {year} {2004}{\natexlab{b}})},\ \Eprint
  {https://arxiv.org/abs/cond-mat/0407066} {cond-mat/0407066} \BibitemShut
  {NoStop}%
\bibitem [{\citenamefont {Bridgeman}\ and\ \citenamefont
  {Chubb}(2017)}]{bridgeman:interpretive-dance}%
  \BibitemOpen
  \bibfield  {author} {\bibinfo {author} {\bibfnamefont {J.~C.}\ \bibnamefont
  {Bridgeman}}\ and\ \bibinfo {author} {\bibfnamefont {C.~T.}\ \bibnamefont
  {Chubb}},\ }\bibfield  {title} {\bibinfo {title} {\emph {Hand-waving and
  Interpretive Dance: An Introductory Course on Tensor Networks}},\ }\href
  {https://doi.org/10.1088/1751-8121/aa6dc3} {\bibfield  {journal} {\bibinfo
  {journal} {J. Phys. A: Math. Theor.}\ }\textbf {\bibinfo {volume} {50}},\
  \bibinfo {pages} {223001} (\bibinfo {year} {2017})},\ \Eprint
  {https://arxiv.org/abs/arXiv:1603.03039} {arXiv:1603.03039} \BibitemShut
  {NoStop}%
\bibitem [{\citenamefont {Perez-Garcia}\ \emph {et~al.}(2008)\citenamefont
  {Perez-Garcia}, \citenamefont {Verstraete}, \citenamefont {Cirac},\ and\
  \citenamefont {Wolf}}]{perez-garcia:parent-ham-2d}%
  \BibitemOpen
  \bibfield  {author} {\bibinfo {author} {\bibfnamefont {D.}~\bibnamefont
  {Perez-Garcia}}, \bibinfo {author} {\bibfnamefont {F.}~\bibnamefont
  {Verstraete}}, \bibinfo {author} {\bibfnamefont {J.~I.}\ \bibnamefont
  {Cirac}},\ and\ \bibinfo {author} {\bibfnamefont {M.~M.}\ \bibnamefont
  {Wolf}},\ }\bibfield  {title} {\bibinfo {title} {\emph {PEPS as unique ground
  states of local Hamiltonians}},\ }\href@noop {} {\bibfield  {journal}
  {\bibinfo  {journal} {Quantum Inf. Comput.}\ }\textbf {\bibinfo {volume}
  {8}},\ \bibinfo {pages} {0650} (\bibinfo {year} {2008})},\ \Eprint
  {https://arxiv.org/abs/arXiv:0707.2260} {arXiv:0707.2260} \BibitemShut
  {NoStop}%
\bibitem [{\citenamefont {Haegeman}\ \emph {et~al.}(2015)\citenamefont
  {Haegeman}, \citenamefont {Zauner}, \citenamefont {Schuch},\ and\
  \citenamefont {Verstraete}}]{haegeman:shadows}%
  \BibitemOpen
  \bibfield  {author} {\bibinfo {author} {\bibfnamefont {J.}~\bibnamefont
  {Haegeman}}, \bibinfo {author} {\bibfnamefont {V.}~\bibnamefont {Zauner}},
  \bibinfo {author} {\bibfnamefont {N.}~\bibnamefont {Schuch}},\ and\ \bibinfo
  {author} {\bibfnamefont {F.}~\bibnamefont {Verstraete}},\ }\bibfield  {title}
  {\bibinfo {title} {\emph {Shadows of anyons and the entanglement structure of
  topological phases}},\ }\href {https://doi.org/doi:10.1038/ncomms9284}
  {\bibfield  {journal} {\bibinfo  {journal} {Nature Comm.}\ }\textbf {\bibinfo
  {volume} {6}},\ \bibinfo {pages} {8284} (\bibinfo {year} {2015})},\ \Eprint
  {https://arxiv.org/abs/arXiv:1410.5443} {arXiv:1410.5443} \BibitemShut
  {NoStop}%
\bibitem [{\citenamefont {{Duivenvoorden}}\ \emph {et~al.}(2017)\citenamefont
  {{Duivenvoorden}}, \citenamefont {{Iqbal}}, \citenamefont {{Haegeman}},
  \citenamefont {{Verstraete}},\ and\ \citenamefont
  {{Schuch}}}]{duivenvoorden:anyon-condensation}%
  \BibitemOpen
  \bibfield  {author} {\bibinfo {author} {\bibfnamefont {K.}~\bibnamefont
  {{Duivenvoorden}}}, \bibinfo {author} {\bibfnamefont {M.}~\bibnamefont
  {{Iqbal}}}, \bibinfo {author} {\bibfnamefont {J.}~\bibnamefont {{Haegeman}}},
  \bibinfo {author} {\bibfnamefont {F.}~\bibnamefont {{Verstraete}}},\ and\
  \bibinfo {author} {\bibfnamefont {N.}~\bibnamefont {{Schuch}}},\ }\bibfield
  {title} {\bibinfo {title} {\emph {{Entanglement phases as holographic duals
  of anyon condensates}}},\ }\href {https://doi.org/10.1103/PhysRevB.95.235119}
  {\bibfield  {journal} {\bibinfo  {journal} {Phys. Rev. B}\ }\textbf {\bibinfo
  {volume} {95}},\ \bibinfo {eid} {235119} (\bibinfo {year} {2017})},\ \Eprint
  {https://arxiv.org/abs/arXiv:1702.08469} {arXiv:1702.08469} \BibitemShut
  {NoStop}%
\bibitem [{\citenamefont {Schuch}\ \emph {et~al.}(2013)\citenamefont {Schuch},
  \citenamefont {Poilblanc}, \citenamefont {Cirac},\ and\ \citenamefont
  {Perez-Garcia}}]{schuch:topo-top}%
  \BibitemOpen
  \bibfield  {author} {\bibinfo {author} {\bibfnamefont {N.}~\bibnamefont
  {Schuch}}, \bibinfo {author} {\bibfnamefont {D.}~\bibnamefont {Poilblanc}},
  \bibinfo {author} {\bibfnamefont {J.~I.}\ \bibnamefont {Cirac}},\ and\
  \bibinfo {author} {\bibfnamefont {D.}~\bibnamefont {Perez-Garcia}},\
  }\bibfield  {title} {\bibinfo {title} {\emph {Topological order in PEPS:
  Transfer operator and boundary Hamiltonians}},\ }\href@noop {} {\bibfield
  {journal} {\bibinfo  {journal} {Phys. Rev. Lett.}\ }\textbf {\bibinfo
  {volume} {111}},\ \bibinfo {pages} {090501} (\bibinfo {year} {2013})},\
  \Eprint {https://arxiv.org/abs/arXiv:1210.5601} {arXiv:1210.5601}
  \BibitemShut {NoStop}%
\bibitem [{\citenamefont {{Molnar}}\ \emph {et~al.}(2018)\citenamefont
  {{Molnar}}, \citenamefont {{Garre-Rubio}}, \citenamefont
  {{P{\'e}rez-Garc{\'\i}a}}, \citenamefont {{Schuch}},\ and\ \citenamefont
  {{Cirac}}}]{molnar:normal-peps-fundamentalthm}%
  \BibitemOpen
  \bibfield  {author} {\bibinfo {author} {\bibfnamefont {A.}~\bibnamefont
  {{Molnar}}}, \bibinfo {author} {\bibfnamefont {J.}~\bibnamefont
  {{Garre-Rubio}}}, \bibinfo {author} {\bibfnamefont {D.}~\bibnamefont
  {{P{\'e}rez-Garc{\'\i}a}}}, \bibinfo {author} {\bibfnamefont
  {N.}~\bibnamefont {{Schuch}}},\ and\ \bibinfo {author} {\bibfnamefont
  {J.~I.}\ \bibnamefont {{Cirac}}},\ }\bibfield  {title} {\bibinfo {title}
  {\emph {{Normal projected entangled pair states generating the same
  state}}},\ }\href {https://doi.org/10.1088/1367-2630/aae9fa} {\bibfield
  {journal} {\bibinfo  {journal} {New J. Phys.}\ }\textbf {\bibinfo {volume}
  {20}},\ \bibinfo {pages} {113017} (\bibinfo {year} {2018})},\ \Eprint
  {https://arxiv.org/abs/arXiv:1804.04964} {arXiv:1804.04964} \BibitemShut
  {NoStop}%
\bibitem [{\citenamefont {Iqbal}\ \emph {et~al.}(2018)\citenamefont {Iqbal},
  \citenamefont {Duivenvoorden},\ and\ \citenamefont
  {Schuch}}]{iqbal:z4-phasetrans}%
  \BibitemOpen
  \bibfield  {author} {\bibinfo {author} {\bibfnamefont {M.}~\bibnamefont
  {Iqbal}}, \bibinfo {author} {\bibfnamefont {K.}~\bibnamefont
  {Duivenvoorden}},\ and\ \bibinfo {author} {\bibfnamefont {N.}~\bibnamefont
  {Schuch}},\ }\bibfield  {title} {\bibinfo {title} {\emph {Study of anyon
  condensation and topological phase transitions from a Z4 topological phase
  using Projected Entangled Pair States}},\ }\href@noop {} {\bibfield
  {journal} {\bibinfo  {journal} {Phys. Rev. B}\ }\textbf {\bibinfo {volume}
  {97}},\ \bibinfo {pages} {195124} (\bibinfo {year} {2018})},\ \Eprint
  {https://arxiv.org/abs/arXiv:1712.04021} {arXiv:1712.04021} \BibitemShut
  {NoStop}%
\bibitem [{\citenamefont {Iqbal}\ \emph
  {et~al.}(2020{\natexlab{a}})\citenamefont {Iqbal}, \citenamefont
  {Casademunt},\ and\ \citenamefont {Schuch}}]{iqbal:rvb-perturb}%
  \BibitemOpen
  \bibfield  {author} {\bibinfo {author} {\bibfnamefont {M.}~\bibnamefont
  {Iqbal}}, \bibinfo {author} {\bibfnamefont {H.}~\bibnamefont {Casademunt}},\
  and\ \bibinfo {author} {\bibfnamefont {N.}~\bibnamefont {Schuch}},\
  }\bibfield  {title} {\bibinfo {title} {\emph {Topological Spin Liquids:
  Robustness under perturbations}},\ }\href@noop {} {\bibfield  {journal}
  {\bibinfo  {journal} {Phys. Rev. B}\ }\textbf {\bibinfo {volume} {101}},\
  \bibinfo {pages} {115101} (\bibinfo {year} {2020}{\natexlab{a}})},\ \Eprint
  {https://arxiv.org/abs/arXiv:1910.06355} {arXiv:1910.06355} \BibitemShut
  {NoStop}%
\bibitem [{\citenamefont {Iqbal}\ \emph
  {et~al.}(2020{\natexlab{b}})\citenamefont {Iqbal}, \citenamefont
  {Poilblanc},\ and\ \citenamefont {Schuch}}]{iqbal:breathing-kagome}%
  \BibitemOpen
  \bibfield  {author} {\bibinfo {author} {\bibfnamefont {M.}~\bibnamefont
  {Iqbal}}, \bibinfo {author} {\bibfnamefont {D.}~\bibnamefont {Poilblanc}},\
  and\ \bibinfo {author} {\bibfnamefont {N.}~\bibnamefont {Schuch}},\
  }\bibfield  {title} {\bibinfo {title} {\emph {Gapped $Z_2$ spin liquid in the
  breathing kagome Heisenberg antiferromagnet}},\ }\href@noop {} {\bibfield
  {journal} {\bibinfo  {journal} {Phys. Rev. B}\ }\textbf {\bibinfo {volume}
  {101}},\ \bibinfo {pages} {155141} (\bibinfo {year} {2020}{\natexlab{b}})},\
  \Eprint {https://arxiv.org/abs/arXiv:1912.08284} {arXiv:1912.08284}
  \BibitemShut {NoStop}%
\bibitem [{\citenamefont {Rams}\ \emph {et~al.}(2018)\citenamefont {Rams},
  \citenamefont {Czarnik},\ and\ \citenamefont
  {Cincio}}]{rams:epsilon-delta-extrapol}%
  \BibitemOpen
  \bibfield  {author} {\bibinfo {author} {\bibfnamefont {M.~M.}\ \bibnamefont
  {Rams}}, \bibinfo {author} {\bibfnamefont {P.}~\bibnamefont {Czarnik}},\ and\
  \bibinfo {author} {\bibfnamefont {L.}~\bibnamefont {Cincio}},\ }\bibfield
  {title} {\bibinfo {title} {\emph {Precise extrapolation of the correlation
  function asymptotics in uniform tensor network states with application to the
  Bose-Hubbard and XXZ models}},\ }\href
  {https://doi.org/10.1103/PhysRevX.8.041033} {\bibfield  {journal} {\bibinfo
  {journal} {Phys. Rev. X}\ }\textbf {\bibinfo {volume} {8}},\ \bibinfo {pages}
  {041033} (\bibinfo {year} {2018})},\ \Eprint
  {https://arxiv.org/abs/arXiv:1801.08554} {arXiv:1801.08554} \BibitemShut
  {NoStop}%
\bibitem [{\citenamefont {Kato}(1966)}]{kato:pert-of-linear-operators}%
  \BibitemOpen
  \bibfield  {author} {\bibinfo {author} {\bibfnamefont {T.}~\bibnamefont
  {Kato}},\ }\href {https://books.google.de/books?id=N\_HysgEACAAJ} {\emph
  {\bibinfo {title} {Perturbation theory for linear operators}}}\ (\bibinfo
  {publisher} {Springer Berlin Heidelberg},\ \bibinfo {year}
  {1966})\BibitemShut {NoStop}%
\bibitem [{\citenamefont {Corboz}\ \emph {et~al.}(2012)\citenamefont {Corboz},
  \citenamefont {Penc}, \citenamefont {Mila},\ and\ \citenamefont
  {Laeuchli}}]{corboz:suN-heisenberg-simplex-solids}%
  \BibitemOpen
  \bibfield  {author} {\bibinfo {author} {\bibfnamefont {P.}~\bibnamefont
  {Corboz}}, \bibinfo {author} {\bibfnamefont {K.}~\bibnamefont {Penc}},
  \bibinfo {author} {\bibfnamefont {F.}~\bibnamefont {Mila}},\ and\ \bibinfo
  {author} {\bibfnamefont {A.~M.}\ \bibnamefont {Laeuchli}},\ }\bibfield
  {title} {\bibinfo {title} {\emph {Simplex solids in SU(N) Heisenberg models
  on the kagome and checkerboard lattices}},\ }\href
  {https://doi.org/10.1103/PhysRevB.86.041106} {\bibfield  {journal} {\bibinfo
  {journal} {Phys. Rev. B}\ }\textbf {\bibinfo {volume} {86}},\ \bibinfo
  {pages} {041106} (\bibinfo {year} {2012})},\ \Eprint
  {https://arxiv.org/abs/arXiv:1204.6682} {arXiv:1204.6682} \BibitemShut
  {NoStop}%
\end{thebibliography}
\end{document}